\documentclass[showpacs,pra,twocolumn,aps,superscriptaddress,floatfix]{revtex4-1}
\usepackage{etoolbox}
\usepackage[caption=false]{subfig}
\newcounter{subeqn} \renewcommand{\thesubeqn}{\theequation\alph{subeqn}}
\newcommand{\subeqn}{
  \refstepcounter{subeqn}
  \tag{\thesubeqn}
}
\makeatother
\usepackage{amsthm}
\usepackage{soul}
\usepackage{xcolor}
\usepackage{bm} \usepackage{amsmath} \usepackage{amssymb} \usepackage{latexsym}
 \usepackage{amsfonts} \usepackage{graphicx}
\renewcommand\Re{\operatorname{Re}}

   \newcommand\romand{\mathrm{d}}
  
\newcommand{\bmH}{\ensuremath{\bar{\mathcal{H}}}}

\newcommand{\eqnrange}[2]{Eqns. (\ref{#1})-(\ref{#2})}

\newcommand{\bthpsi}{\ensuremath{\bar{\delta \hat{\Psi}}}}
\newcommand{\bthpsip}{\ensuremath{\bar{\delta \hat{\Psi}}_{\perp}}}
\newcommand{\bvphi}{\ensuremath{\bar{\varphi}}}
\newcommand{\bM}{\ensuremath{\mathbf{M}}}
\newcommand{\bG}{\ensuremath{\mathbf{G}}}
\newcommand{\bC}{\ensuremath{\mathbf{C}}}
\newcommand{\dalp}{\ensuremath{\delta \hat{a}}}
\newcommand{\dalpd}{\ensuremath{\delta \hat{a}^{\dagger}}}

\newcommand{\ba}{\begin{align}}
\newcommand{\ea}{\end{align}} 
\newcommand{\dc}{\ensuremath{\delta \hat{c}}}
\newcommand{\db}{\ensuremath{\delta \hat{b}}}
\newcommand{\eea}{\end{eqnarray}}
       
\usepackage{color}
\newcommand{\ha}{\ensuremath{\hat{a}}}
\newcommand{\had}{\ensuremath{\hat{a}^{\dagger}}}
\newcommand{\hpsi}{\ensuremath{\hat{\Psi}}}
\newcommand{\hpsid}{\ensuremath{\hat{\Psi}^{\dagger}}}

\newcommand{\hxi}{\ensuremath{\hat{\xi}}}
\newcommand{\hxid}{\ensuremath{\hat{\xi}^{\dagger}}}

\newcommand{\hd}{\ensuremath{\hat{d}}}
\newcommand{\hdd}{\ensuremath{\hat{d}^{\dagger}}}

\newcommand{\hP}{\ensuremath{\hat{P}}}
\newcommand{\hZ}{\ensuremath{\hat{Z}}}
\newcommand{\dceff}{\ensuremath{\Delta_{\mathrm{c}}^{\mathrm{eff}}}}
\newcommand{\aout}{\hat{a}_{\mathrm{out}}}
\newcommand{\aoutd}{\hat{a}_{\mathrm{out}}^{\dagger}}
\newcommand{\ain}{\hat{a}_{\mathrm{in}}}

\newcommand{\iout}{\hat{I}_{\mathrm{out}}}

\newcommand{\hR}{\ensuremath{\hat{R}}}

\newcommand{\bL}{\ensuremath{\mathbf{\Lambda}}}
\soulregister{\cite}{1}
\soulregister{\eqnref}{1}
\usepackage{placeins}

\newcommand{\eqnref}[1]{Eq.~(\ref{#1})}
\newcommand{\figref}[1]{Fig.~\ref{#1}}
\makeatletter

\newcommand{\Rmnum}[1]{\expandafter\@slowromancap\romannumeral #1@}
\makeatother

\begin{document}
\title{Bloch Oscillations of Cold Atoms in a Cavity: Effects of Quantum Noise}

\author{B.\ Prasanna Venkatesh}
\affiliation{Department of Physics and Astronomy, McMaster University, 1280 Main
St.\ W., Hamilton, ON, L8S 4M1, Canada} 
\author{D.\ H.\ J.\ O'Dell}
\affiliation{Department of Physics and Astronomy, McMaster University, 1280 Main
St.\ W., Hamilton, ON, L8S 4M1, Canada}
\begin{abstract}
We extend our theory of Bloch oscillations of cold atoms inside an optical cavity [B. P. Venkatesh
\textit{et al.}, Phys. Rev. A \textbf{80}, 063834 (2009)] to include the effects of quantum noise arising from coupling to external modes. The noise acts as a form of quantum measurement backaction by perturbing the coupled dynamics of the
atoms and the light. We take it into account by solving the Heisenberg-Langevin equations for
linearized fluctuations about the atomic  and optical meanfields and examine how this influences the
signal-to-noise ratio of a measurement of external forces using this system. In particular, we
investigate the effects of changing the number of atoms, the intracavity lattice depth, and the
atom-light coupling strength, and show how resonances between the Bloch oscillation dynamics and the
quasiparticle spectrum have a strong influence on the signal-to-noise ratio as well as heating effects. One of the hurdles we overcome in this paper is the proper treatment of fluctuations about time-dependent meanfields in the context of cold atom cavity-QED . 
\end{abstract}
\maketitle

\section{Introduction}\label{sec1}
When quantum particles in a periodic potential of period $d$ are subject to a weak additional
constant force $F$ they do not uniformly accelerate like free particles, but instead undergo Bloch
oscillations \cite{bloch28} at an angular frequency given by:
\begin{align}
\omega_{\mathrm{B}} = Fd/ \hbar  \ . \label{eq:blochfrequency}
\end{align}
Bloch oscillations (BOs) of cold atoms in optical lattices were first observed in 1996 by uniformly
accelerating the lattice \cite{salomon96}: in a frame co-moving with the lattice the atoms
experience a constant force. At about the same time, the accelerating lattice method was used to
observe Wannier-Stark ladders \cite{raizen96}, which are a different aspect of the same ``tilted
lattice'' physics. The method has subsequently been employed to realize beam splitters for atom
optics capable of large momentum transfers, see, e.g.\ \cite{battesti04}.

In gravity-driven BOs the lattice is held fixed in space but oriented vertically so that gravity
provides the force $F_{g}=mg$  on the atoms (of mass $m$). From Eq.\ (\ref{eq:blochfrequency}), a
measurement of $\omega_{\mathrm{B}}$ corresponds to a measurement of the applied force $F$ if we
know $d/\hbar$. This Bloch oscillator may be viewed as an interferometer in momentum space
\cite{carusotto05} and has been experimentally demonstrated by a number of groups
\cite{anderson98,morsch01,roati04,ferrari06}.  For example, the experiment \cite{ferrari06} used
gravity-driven BOs of strontium atoms to measure the local acceleration due to gravity at the level
of $\Delta g/g=5 \times 10^{-6}$. Like any interferometer, long coherence times are crucial for
precision measurements and in \cite{ferrari06} the BOs were coherent over 7 s, corresponding to
$\approx 4000$ oscillations. This remarkable degree of coherence was greatly facilitated by the
choice of strontium atoms, which have very weak $s$-wave scattering, and thus dynamical
instabilities normally associated with superflow in lattices \cite{burger01,wu03} were highly
suppressed. Variations on this scheme that improve the visibility of the BOs, including frequency
\cite{ivanov08} and amplitude \cite{poli11,tarallo12} modulation of the lattice, have allowed for
the measurement of gravity at the level of $\Delta g/g=10^{-9}$. In these latest experiments the BOs
were coherent for over 20 seconds.

The experiments referred to above all involve destructive measurements of the BOs  due to the nature
of the imaging process of the atoms, whether it be \emph{in situ} or by a time-of-flight technique
after the lattice has been switched off \cite{tarallo12}. Therefore, a precision measurement of
$\omega_{\mathrm{B}}$ by the above methods  requires that the experiment be re-run many times, each
run being for a slightly different hold time, so that the oscillations can be accurately mapped out.
This not only takes a long time, but also requires that the initial conditions be recreated as
faithfully as possible for each run.

\begin{figure}
\includegraphics[height = 9cm]{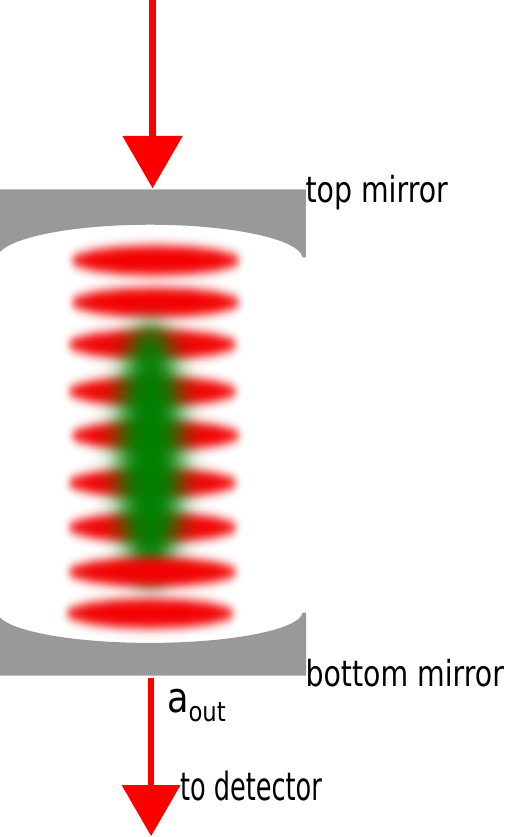}
\caption{(Color online) Schematic of the precision measurement proposal in \cite{bpvenkatesh09}. A
dilute cloud of cold atoms undergoes BOs
in the combined intracavity lattice potential and the acceleration due to gravity. The transmitted
light field's intensity and phase are
modulated at the Bloch frequency. A in-situ precise measurement of the Bloch frequency (and hence
the force) can be performed by detecting
the transmitted light. }
\label{fig:schemepic} 
\end{figure}

In \cite{bpvenkatesh09} we proposed a scheme for continuous  (i.e.\ non-destructive) measurements of
BOs based upon placing the atoms inside a Fabry-Perot optical resonator which would allow for an
estimate of $\omega_{\mathrm{B}}$ from the data acquired over a single run. A related scheme has
also been independently proposed for ring cavities \cite{peden09}. The periodic potential is now
provided by the standing wave of light which forms inside the cavity when it is pumped by a laser.
Orienting the cavity vertically, the atoms execute BOs along the cavity axis as depicted in Fig.\
\ref{fig:schemepic}.  The enhanced atom-light coupling inside a high-Q cavity means that the
oscillating atoms imprint a detectable periodic modulation on both the phase and amplitude of the
light which can be seen either in transmission or reflection. Thus, the measurement is performed
upon the light leaking out of the cavity rather than directly upon the atoms.

The strong atom-light coupling that can be realized in cavity-QED stands in contrast to the case of
optical lattices in free space where the atoms exert only a tiny backaction upon the light. The
optical dipole interaction between a single cavity photon and a single atom is characterized by the
Rabi frequency $g_0= (\mu/\hbar) \sqrt{\hbar \omega_{c}/(\varepsilon_{0} V)}$, where
$\omega_{\mathrm{c}}$ and $V$ are the frequency and volume of the relevant cavity mode and $\mu$ is
the atomic transition dipole moment. Defining the cooperativity $C \equiv g_{0}^2/(2 \gamma
\kappa)$, where $2 \gamma$ is the spontaneous emission rate of the atom in free space and $2 \kappa$
is the energy damping rate of the cavity, $1/C$ is the number of atoms required to strongly perturb
the light field. The normal mode splitting that results from strong coupling has been directly
observed in a number of cold atom optical cavity experiments
\cite{boca04,maunz05,klinner06,colombe07}. In the experiment \cite{colombe07}, which was performed
with a Bose-Einstein condensate, the cooperativity was $C=145$. Even more pertinently, these
systems have been used to detect the presence of single atoms
\cite{mabuchi96,munstermann99,trupke07}, as well as to follow their dynamics in real time
\cite{ye99,pinske00}.  The collective dynamics of ultracold atomic gases have also been tracked
using cavities \cite{nagorny03,gupta07,brennecke08}.  The key experimental steps necessary for the
continuous monitoring of BOs in a cavity have, therefore, already been demonstrated.

The drawback with any continuous measurement scheme is measurement backaction. In cavities this
backaction typically takes the form of cavity photon number fluctuations which lead to random force
fluctuations on the atoms, as is evident in the erratic nature of the single atom trajectories seen
in the experiments \cite{ye99,pinske00} referred to above. In the many atom context, quantum
measurement backaction generally manifests itself in a heating of the atom cloud (although under
some circumstances it can lead to cooling \cite{horak97}). In the cavity-optomechanical regime
(where the collective motion can be modelled as a harmonic oscillator of angular frequency $\omega$)
the heating rate is expected to be $R=(x_{\mathrm{zpf}}/\hbar)^2 S_{\mathcal{FF}}(-\omega)$
\cite{marquardt07}, where $x_{\mathrm{zpf}}$ is the zero-point fluctuation and $S_{\mathcal{FF}}$ is
the spectral density of the force fluctuations (which is directly proportional to the cavity photon
number fluctuations). This heating rate is in agreement with observations when convolved with
technical fluctuations \cite{murch08}.

In the system considered in this paper (see Fig.\ \ref{fig:schemepic}), we can divide the backaction
into two types. One type comes from the fact that the atoms sit in an optical lattice whose depth is
periodically modulated in time at the frequency $\omega_{\mathrm{B}}$ due to the effect of BOs. This
backaction is a classical effect in the sense that it occurs even when the light field is treated
classically (no photons). The nonlinearity that arises from this backaction can lead to swallowtail
loops in the atomic band structure \cite{bpvenkatesh11,coles12} that mimics the effects of direct
atom-atom interactions \cite{wu01,diakonov02,mueller02,smerzi02}. These loops are the counterpart in
the atomic wave function of optical bistability in the light
\cite{nagorny03,gupta07,brennecke08,chen10}.   The second type of backaction arises only when the
fluctuations due to the discrete photon nature of the light field are taken into account and is
related to the heating effect mentioned above. The characteristic frequency of these latter
fluctuations is $\kappa$ which is much larger than $\omega_{\mathrm{B}}$.  

The first type of backaction was analyzed in our previous paper \cite{bpvenkatesh09} where our main
aim was to show that, despite the self-generated time modulation of the intracavity optical lattice,
the Bloch acceleration theorem still applies and the  BO frequency is not modified (although
harmonics can be generated). This latter result is clearly very important if the cavity BO method is
to be used for precision measurements and may be viewed as a consequence of the fact that the
formula (\ref{eq:blochfrequency}) does not depend on the depth of the lattice, only its spatial
period. An estimate of the effects of the second type of backaction was also given in our previous
paper, but this estimate was obtained under the assumption that the photon number fluctuations were
purely due to the photon shot noise found in a coherent state of light. This ignores the
correlations that build up between the atoms and the light inside the cavity and our main aim in
this paper is to solve the dynamics of the coupled photon and atom fluctuations systematically from
first principles and thereby capture these correlations. This will allow us to properly determine
the sensitivity of the measurement of the Bloch frequency to quantum fluctuations.

The plan of this paper is as follows: in Section \ref{sec:hamiltonian} we introduce the physical
system, the associated hamiltonian, and the equations of motion. We then review in Sections
\ref{sec:meanfield} and \ref{sec:mftresults} the meanfield approximation and the associated
numerical results which were the focus of our previous paper \cite{bpvenkatesh09},  before
introducing in Section \ref{sec:fluctlin} the main model to be treated in this paper which adds
quantum fluctuations. This is an elaboration of the linearization approach presented in, e.g.\
\cite{horak01,szirmai09,szirmai10}, to include a time-dependent meanfield component (due to the
BOs). The fluctuations correspond to quasiparticles (excitations out of the meanfield), and their
spectrum is analyzed in Section \ref{sec:excspect} and then used to help interpret the numerical
results for the quantum dynamics presented in Section \ref{sec:fluctdyn}. We also develop a simple
rate equation picture, valid in the weak coupling regime,  to help us understand the rate of
quasiparticle excitation. Following this we change gears slightly and apply the above results to
investigate how quantum fluctuations affect a precision measurement of $\omega_{\mathrm{B}}$ by
calculating the signal-to-noise ratio (SNR). We present the theory lying behind these
calculations in Section \ref{sec:snr} and in Section \ref{sec:snrresults} we examine the results, 
paying particular attention to whether or not there is an optimal value for the atom-light coupling
parameter $\beta = NU_0/\kappa$. We also present results illustrating the dependence of the SNR on
other system parameters such as the number of atoms and the intracavity lattice depth.  We summarize
our results and give some further perspective in Section \ref{sec:conclusion}. We have also provided three
appendices that give details omitted from the main text: the first derives an approximation wherein the cavity field is assumed to be in a coherent state and the atomic fluctuations about the meanfield are treated as independent oscillators, the second discusses the effects that BOs have on cavity cooling, and the third discusses our approach to calculating two-time correlation functions.

\section{Hamiltonian and Equations of Motion}
\label{sec:hamiltonian}

Our system consists of a gas of $N$ bosonic atoms inside a vertically oriented Fabry-Perot optical
cavity. A single cavity mode of frequency $\omega_{\mathrm{c}}$ is  coherently pumped by a
laser with frequency $\omega_{\mathrm{p}}$ that is detuned from both the atomic and the
cavity resonance frequencies. This sets up a standing wave mode along the cavity axis of the form
$\cos(k_{\mathrm{c}} z)$, where $k_{\mathrm{c}} = \omega_{\mathrm{c}}/c$. The relevant frequency
relations are characterized by the two detunings
    \begin{align}
    \Delta_{\mathrm{c}} & \equiv  \omega_{\mathrm{p}}-\omega_{\mathrm{c}}, \subeqn
\refstepcounter{equation}  \\
    \Delta_{a} & \equiv  \omega_{\mathrm{p}}-\omega_{a}, \subeqn
    \end{align}
where $\omega_{a}$ is the atomic transition frequency. In the dispersive regime, the occupation of
the excited atomic state is vanishingly
small and it can be adiabatically eliminated. A one-dimensional hamiltonian for the atom-cavity
system in the dispersive regime can then be written as \cite{maschler05,larson08} 
    \begin{align}
    \hat{H} &= -\hbar \Delta_{\mathrm{c}} \ha^{\dagger} \ha + i \hbar \eta \left (\ha^{\dagger} -
    \ha\right) \nonumber \\
    & + \int \romand z \ \hpsi^{\dagger} \left[-\frac{\hbar^2}{2M} \frac{\partial^2}{\partial z^2} +
    \hbar U_0    \ha^{\dagger} \ha \cos^2 (k_{\mathrm{c}}z) -Fz \right] \hpsi,
    \label{eq:hamiltonian}
    \end{align}
where $\hpsi(z,t)$ and $\ha(t)$ are the field operators for the atoms and the cavity photons which
obey the equal time bosonic commutation relations 
$[\hat{\Psi}(x,t),\hat{\Psi}^{\dag}(x',t)]=\delta(x-x')$, and $[\hat{a}(t),\hat{a}^{\dag}(t)]=1$,
respectively. The single atom dispersive light shift has been denoted by $U_0 \equiv
g_{0}^2/\Delta_a$.

The hamiltonian has been written in a frame rotating with the pump laser frequency
$\omega_{\mathrm{p}}$, and this leads to the appearance of the two detunings. The first term is just
the free evolution of the cavity mode. The second term represents the laser coherently pumping the
cavity at rate $\eta$, and  the third term describes the atomic part of the hamiltonian. The first
two terms of the atomic part represent the  kinetic energy  and a light induced potential energy.
This latter term can either be understood as the atom moving in a periodic potential with average
amplitude $\hbar U_0 \langle \ha^{\dagger}\ha \rangle$ or, if combined with the first term in the
hamiltonian, as a shift in the resonance frequency of the cavity due to the coupling between the
atom and the field. The third term in the atomic part provides the external force that drives the
BOs. We assume this force arises from the vertical orientation ($z$ increases in the downward
direction) of the cavity and is given by $F = Mg$.

We have not included direct atom-atom interactions in the hamiltonian (\ref{eq:hamiltonian}) 
because under realistic experimental conditions they are three orders of magnitude smaller than the recoil energy $E_{\mathrm{R}} \equiv \hbar^2 k_{\mathrm{c}}^2/(2M)$ which characterizes the single-particle energy (kinetic and potential) of an atom in an optical lattice. Consider, for example, the meanfield interaction energy per particle $E_{\mathrm{int}}/N= (2 \pi \hbar^2 a_{s} N/M ) \int \vert \Phi(\mathbf{r}) \vert^4 d^{3}r$ for a cloud of $N=5 \times 10^4$ $^{87}$Rb atoms trapped in a 178 $\mu$m long cavity \cite{brennecke08}. Here $a_{s}=5.3$ nm is the $s$-wave scattering length. We take the normalized 3D wave function $\Phi(\mathbf{r})$ to be the product of a ground band Bloch wave that extends 178 $\mu$m along $z$ and a gaussian 25 $\mu$m wide in the transverse plane. Then, evaluating the Bloch wave for a lattice which is 3 $E_{\mathrm{R}}$ deep and made from $780$ nm light (456 wells are occupied), we find the ratio $E_{\mathrm{int}}/N : E_{\mathrm{R}} = 1.1 \times 10^{-3}$. The interactions can be tuned to smaller values still using a Feshbach resonance: the experiment \cite{gustavsson08} increased the dephasing time of BOs from a few oscillations to 20,000 using this technique. The fact that the atoms all interact with a common light field whose magnitude is modified by the sum of their individual couplings gives rise to a nonlinearity (the classical backaction referred to above) that is in some ways analogous to that due to direct interactions \cite{bpvenkatesh11,zhou09}, but in other ways differs and can lead to novel behavior \cite{maschler05,larson08,mottl12}. 

Natural units for the length and energy in cavity-QED are given by $1/k_{\mathrm{c}}$ and the recoil
energy $E_{\mathrm{R}}$, respectively. From here on we scale all
lengths by $1/k_{\mathrm{c}}$ and consequently define $x \equiv k_{\mathrm{c}} z $. We scale
frequencies by the recoil frequency $\omega_{\mathrm{R}} \equiv E_{\mathrm{R}}/\hbar$ and time by
$1/\omega_{\mathrm{R}}$ and retain the same symbols for the scaled variables. The
Heisenberg-Langevin equations of motion for the light and atomic field operators in the scaled
variables are \cite{maschler05}
\begin{align}
 i\frac{d \ha}{dt} = \Big[ -\Delta_{\mathrm{c}}+  \int  dx \ \hpsid(x,t)  & \hpsi(x,t)     U_0 
\cos^2(x) -i\kappa \Big] \ha 
\nonumber\\
&+ i\eta + i \sqrt{2 \kappa} \hxi(t) \refstepcounter{equation} \subeqn \label{eq:HLlight} 
\setcounter{subeqn}{1} 
 \\
i \frac{\partial \hpsi}{\partial t} =    \Big[ -\frac{\partial^2}{\partial x^2} +   U_0 \had \ha
\cos^2 & (x)   - fx \Big] \hpsi 
\subeqn \label{eq:HLatom}
\end{align}
where $f \equiv F/(\hbar k_{\mathrm{c}} \omega_{\mathrm{R}}) = \omega_{\mathrm{B}}/(\pi
\omega_{\mathrm{R}})$ is the dimensionless form for
the external force. The operator $\hxi(t)$ is the Langevin term and is assumed to be Gaussian white
noise with the only non-zero correlation
being
\begin{align}
 \langle \hxi(t) \hxid (t')\rangle =  \delta (t-t'). \subeqn \label{eq:whitenoise} 
\end{align}
Mathematically, the Langevin noise terms are necessary in order to preserve the commutation relation
$[\hat{a}(t),\hat{a}^{\dag}(t)]=1$  in an open system. Physically, their origin is vacuum
fluctuations of the electromagnetic field that are transmitted into the cavity via the mirrors and
they thus only appear in the equations for the light field. Nevertheless, the noise is conveyed to
the atomic dynamics by the atom-light coupling.

\section{Meanfield dynamics: theory}
\label{sec:meanfield}

The approach we follow in this paper is based upon a separation of the field operators into
meanfield and quantum parts:
\begin{align}
\ha(t) &=  \alpha(t) + \dalp(t)    \subeqn  \label{eq:photsplit} \refstepcounter{equation} 
\setcounter{subeqn}{1}   \\
 \hpsi(x,t) &= \sqrt{N} \varphi(x,t) +  \delta \hpsi(x,t) \subeqn \label{eq:atomsplit}.
\end{align}
In the meanfield approximation the light is assumed to be in a classical state with amplitude
$\alpha(t) = \langle \ha(t) \rangle$, where $\vert \alpha(t) \vert^2$ corresponds to the average
number of photons in the cavity, and the atoms are assumed to all share the same single-particle
wave function $\varphi(x,t) = \langle \hpsi (x,t) \rangle / \sqrt{N}$. The equations of motion for
the meanfield amplitudes $\alpha(t)$ and $\varphi(x,t)$ are
\begin{align}
\refstepcounter{equation}
i \frac{d \alpha(t)}{dt} &= \left[-\Delta_{\mathrm{c}} + N U_0 \langle \cos^2(x) \rangle - i\kappa
\right] \alpha(t) + i \eta
\label{eq:mftlight} \subeqn  \setcounter{subeqn}{1}   \\
i \frac{\partial \varphi(x,t)}{\partial t} &= \left[-\frac{\partial^2}{\partial x^2} +\vert
\alpha(t) \vert^2 U_0 \cos^2(x)
- fx  \right] \varphi(x,t) \label{eq:mftatom} \subeqn
\end{align}
where the second equation has the form of a Schr\"{o}dinger equation. The expectation value
\begin{equation}
\langle \cos^2(x) \rangle(t) = \int dx \vert \varphi(x,t) \vert^2 \cos^2(x) \subeqn 
\label{eq:couplingintegral}
\end{equation} 
that appears in the first of these equations provides the time-dependent coupling between the atomic
probability density and the cavity mode function. Multiplying this integral is the collective
atom-cavity coupling parameter $NU_{0}$.  When measured in units of the cavity linewidth we denote
this parameter by $\beta$ 
\begin{equation}
\beta \equiv NU_0/\kappa \, .
\label{eq:betadefinition}
\end{equation}
We illustrate the effect that $\beta$ has on the  meanfield dynamics in Figs.\
\ref{fig:mftlatplotdc1_changecoup} and \ref{fig:latticedepth&FT} below.

In reference \cite{bpvenkatesh11} we studied the influence the classical backaction nonlinearity has
upon the band structure of atom-cavity systems. The band structure is given by the steady state
solutions [$\dot{\alpha}=0$, $\varphi(x,t)=\varphi(x) \exp (-i \mu t/\hbar)$] of the coupled
equations of motion (\ref{eq:mftlight}) and (\ref{eq:mftatom}) in the absence of the external force
$f$. It is straightforward to see that, despite the nonlinearity, exact solutions of the steady
state problem are given by Mathieu functions (like in the linear problem of a quantum particle in a
fixed cosine potential). Mathieu functions are Bloch waves and so can be labelled by a band index
$b$ and quasimomentum $q$ \cite{A+Snew}
\begin{align}
\varphi_{q,b}(x) = e^{iqx} \mathcal{U}_{q,b}(x) \label{eq:blochfn}
\end{align}
where $\mathcal{U}_{q,b}(x+\pi) = \mathcal{U}_{q,b}(x)$ has the same period as the lattice. In the
reduced zone picture $q$ is restricted to lie in the first Brillouin zone $-1 < q \le 1$.
Substituting the Bloch wave solution into the equations of motion yields the steady state equations
\begin{align}
\refstepcounter{equation}
\alpha_{ss} &= \frac{i \eta}{\Delta_{\mathrm{c}}-N U_0 \langle \cos^2(x)  \rangle + i \kappa }
\subeqn \label{eq:lightsteady} 
\setcounter{subeqn}{1} 
\\
\mu_{q,b} \, \mathcal{U}_{q,b}(x) &=  \left[\left(-i \frac{\partial}{\partial x}+q\right)^2 +\vert
\alpha_{ss} \vert^2 U_0 \cos^2(x) \right]
\mathcal{U}_{q,b}(x) \subeqn
\label{eq:atomsteady}
\end{align}
where the subscript $ss$ denotes ``steady state''. Solving these equations one obtains a \emph{band
structure} analogous to that in the linear case but with the striking difference that the
nonlinearity can lead to swallowtail loops in the bands. It is important to appreciate that this
band structure is not for the atoms alone, but for the combined atom-cavity system.  For example,
the eigenvalue $\mu$ is actually a chemical potential rather than the band energy (for the
underlying energy functional with the light adiabatically eliminated see \cite{bpvenkatesh11}), and
another difference from the linear case is that the lattice depth $s=U_{0} \vert \alpha_{ss}
\vert^2$ is not fixed,  but instead depends on the values of $\{b,q\}$. So, for example, the lattice
depth changes during a BO as $q$ is swept along the band.

The external force $f$ breaks the spatial periodicity and means that Bloch waves are replaced by
Wannier-Stark states as the stationary solutions of the equations of motion (in fact, in
finite systems the Wannier-Stark states are resonances rather than true eigenstates \cite{gluck02}).
The spatial periodicity can be restored by applying the unitary transformation $\bvphi(x,t)
=\exp(-iftx) \varphi(x,t)$ which removes the $fx$ term  appearing in the hamiltonian in the
Schr\"{o}dinger equation (\ref{eq:mftatom}) and introduces a shift $ft$ into the momentum operator 
\begin{align}
\mathcal{H}=-\frac{\partial^2}{\partial x^2} & +   s(t)  \cos^2(x)  -fx \nonumber \\
  \longrightarrow  \bar{\mathcal{H}} = & \left( -i \frac{\partial}{\partial x} + ft \right)^2 + s(t)
\cos^2(x) \ . \label{eq:spHtrans}
\end{align}
We denote the frame resulting from this transformation as the transformed frame (TF), and the
original frame as the lab frame (LF).

Let us now consider the dynamics  under the influence of the force term. We take the initial atomic
state $\bvphi(x,t=0)  = \varphi(x,t=0)$ to be a Bloch state in the ground band with quasimomentum
$q=q_{0}$. In the adiabatic approximation the atoms remain in the ground band but the force causes the quasimomentum to sweep
periodically through the first Brillouin zone in accordance with the Bloch acceleration theorem
\begin{align}
 q(t) = q_0 + ft \label{eq:blochaccthm}
\end{align}
as can be seen by comparing Eqns.\ (\ref{eq:atomsteady}) and (\ref{eq:spHtrans}). In fact, a careful
analysis \cite{kroemer86} shows that Eq.\ (\ref{eq:blochaccthm}) holds even when adiabaticity is
broken and interband transitions are allowed providing these transitions are ``vertical'', i.e. they
conserve $q$.

This standard approach to BOs remains valid even when the lattice depth is modulated in time, as
takes place in cavities, because amplitude modulation does not break the spatial periodicity of the
potential and so cannot change $q$ \cite{bpvenkatesh09}. We therefore find that at any later time
$t$, the \emph{exact} atomic meanfield can be expressed as  
\begin{equation}
\varphi(x,t) = \exp \left[i(q_0+ft)x \right] \mathcal{U}(t) \ .
\end{equation}
In general $\mathcal{U}(t)$ is in a superposition of bands and so is no longer the steady state
solution of Eqns.\ (\ref{eq:lightsteady}) and (\ref{eq:atomsteady}), although it does retain its
Bloch form.  The advantage of the TF is that the quasimomentum is frozen at its initial value and we
have 
\begin{equation}
 \bvphi(x,t) = \exp \left[iq_{0} x\right] \mathcal{U}(t) 
\end{equation}
so that it is only the spatially periodic function $\mathcal{U}(t)$ that evolves in time. From the
point of view of numerical computation this allows us to work with a basis of periodic functions (we
normalize our wave functions over one period of the lattice). At any given time a relatively small
number of basis functions can accurately describe the atomic meanfield state and this greatly
reduces the numerical effort in the calculation of BOs.

By working in terms of Bloch waves, our approach is predisposed towards
treating wave functions which are localized in momentum space rather than coordinate space. This
choice is sensible because momentum space is a natural setting for BOs as is evident from
\eqnref{eq:blochaccthm}. This is also in line with existing experiments demonstrating cold atom BOs
in free space optical lattices
\cite{salomon96,anderson98,morsch01,roati04,ferrari06,ivanov08,poli11}, where the initial state is
generally a fairly narrow wavepacket in momentum space. In this paper we shall therefore restrict
ourselves to states that are completely localised in quasimomentum ($\delta$-function wave packet).

\begin{figure}
\includegraphics[width=0.9\columnwidth]{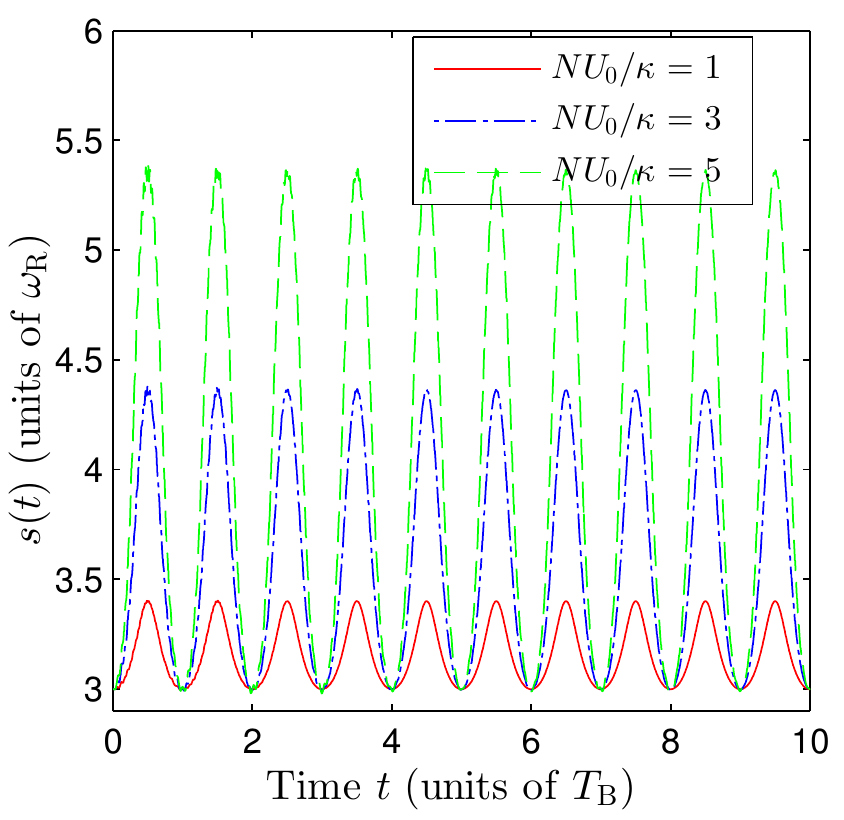}
\caption{(Color online) Intracavity optical lattice depth $s(t) \equiv U_0 \vert \alpha(t) \vert^2$
in units of the atomic recoil frequency
$\omega_{R}$ plotted as a function of time. The curves, which are each for a different value of the
collective atom-cavity coupling
parameter $NU_0/\kappa$, were obtained by solving the meanfield equations of motion Eqns.\
(\ref{eq:mftlight}) and (\ref{eq:mftatom}) and
illustrate the fact that the change in lattice depth over one Bloch oscillation increases with
$NU_0/\kappa$. In order to maintain a minimum
lattice depth of $3 E_{\mathrm{R}}$ as  $NU_0/\kappa$ was
increased by changing $U_0 = \{1,3,5 \} u_0$, where $u_0 =   7 \times 10^{-3}
\omega_{\mathrm{R}}$,  we also changed the pumping strength
as $\eta = \{ 30.7,24.2,24.3\} \kappa$, giving mean photon numbers $\{458,172,117 \}$, respectively.
The other parameter values used in this plot are
$\Delta_{\mathrm{c}} = -0.75 \, \kappa$, $\kappa =345 \, \omega_{\mathrm{R}}$, and $N = 5 \times
10^4$. For all the plots in this paper the
force is such that the Bloch frequency has the value $\omega_{\mathrm{B}} =
\omega_{\mathrm{R}}/4$.}
\label{fig:mftlatplotdc1_changecoup}
\end{figure}
\begin{figure*}
\centering
  \subfloat[Lattice depth as a function of time 
]{\label{fig:mftlatatsnrdip}\includegraphics[width=.45\textwidth]{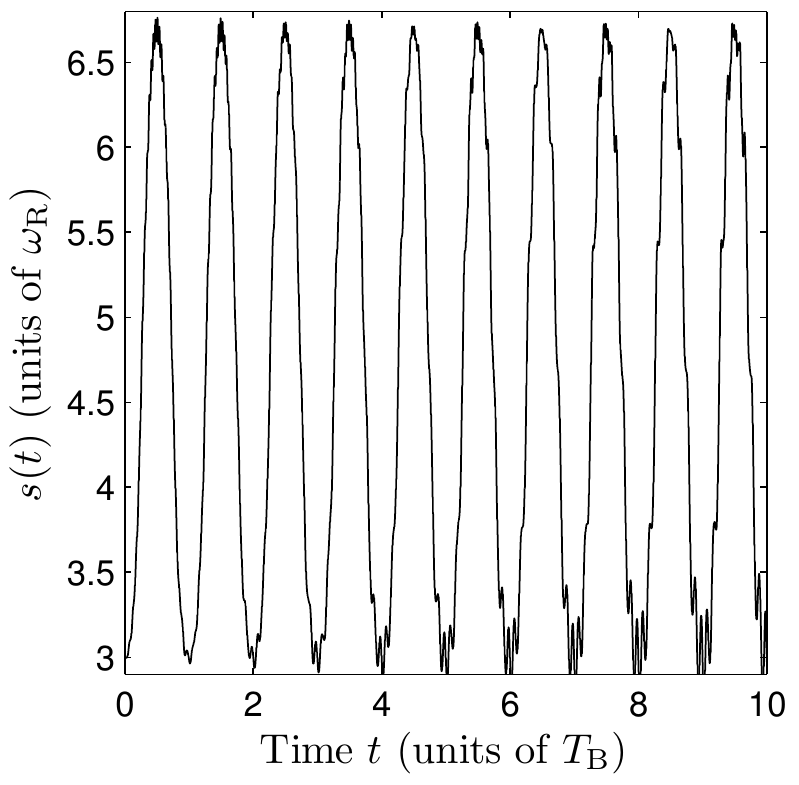}}
  \hfill
 \subfloat[Fourier transform in time of lattice depth]{\label{fig:mftlatFTatsnrdip}
\includegraphics[width=.45\textwidth]{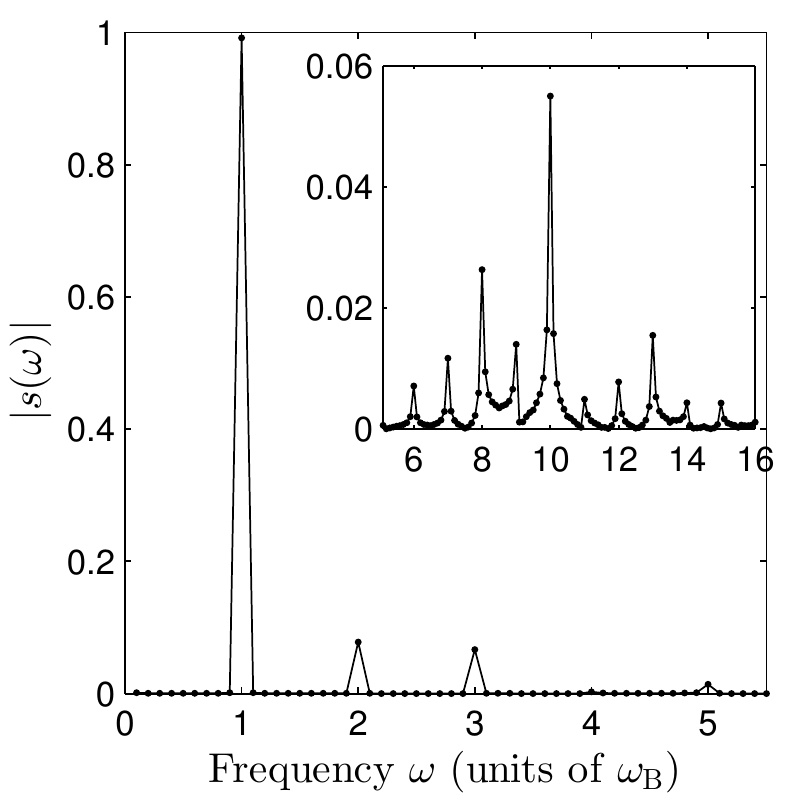}}
\caption{The lattice depth $s(t)$ in units of the atomic recoil frequency $\omega_{\mathrm{R}}$ is
shown in (a) and its Fourier transform
$s(\omega)$ is given in (b). We have increased the atom-cavity coupling from Fig.\
\ref{fig:mftlatplotdc1_changecoup} to $NU_{0}/\kappa =
7.75$. At this larger value some fast fluctuations on top of the slow BO become visible. Their
frequency is dominated by a harmonic at $10
\, \omega_{B}$ as can be seen in the inset.}
\label{fig:latticedepth&FT}
\end{figure*} 

\section{Meanfield dynamics: results}
\label{sec:mftresults}

We now present our numerical results for the meanfield dynamics. The initial state at time $t=0$ is
taken to have quasimomentum $q=0$, and be given by the solutions $\alpha_{ss}$ and
$\mathcal{U}_{0,0}(x)$ of the meanfield steady state equations [Eqns.\ (\ref{eq:lightsteady}) and
(\ref{eq:atomsteady})] for atoms in the ground band. This state is propagated in time using the
meanfield equations of motion [\eqnref{eq:mftlight} and \eqnref{eq:mftatom}]. The reasons for our choices for the
parameter values $\{U_{0},N,\eta,\Delta_{c},\kappa \}$ will be explained at the end of this Section.

Under the action of the external force the atoms begin performing BOs, which for atoms in extended
Bloch states gives rise to a breathing motion of the atomic density distribution on each lattice
site \cite{bpvenkatesh09}. The classical backaction  imprints an
oscillation on the amplitude and phase of the light field at the Bloch frequency
$\omega_{\mathrm{B}}$. In \figref{fig:mftlatplotdc1_changecoup} we plot the time-dependence of the intracavity lattice depth $s(t) = U_0 \vert \alpha(t) \vert^2$ seen by atoms, which is proportional to the number of cavity photons $\vert \alpha
\vert^{2}$.  The experimental signature of the BOs is the photon current transmitted by the cavity,
and this is given in the meanfield approximation by $\kappa \vert \alpha(t) \vert^2$, and hence is
directly proportional to $s(t)$.

The size of the backaction is controlled by the collective coupling $\beta = N U_{0}/ \kappa$, as is
apparent from the different curves in \figref{fig:mftlatplotdc1_changecoup}. As $\beta$ is increased
the change in the lattice depth over a Bloch period increases and hence the visibility or
\emph{contrast} of the BOs as measured by a photon detector outside the cavity increases also.  We
define the contrast $\epsilon$ as 
\begin{align}
 \epsilon \equiv \left( s_{\mathrm{max}} - s_{\mathrm{min}} \right)/\left( s_{\mathrm{max}}+ 
s_{\mathrm{min}} \right) .
\label{eq:contrastdef}
\end{align} 
Each curve in \figref{fig:mftlatplotdc1_changecoup} has a
different pumping strength $\eta$ in order to maintain the same minimum lattice depth of $3
E_{\mathrm{R}}$. If the lattice becomes too shallow interband transition rates (e.g.\ due to Landau-Zener tunnelling around the band edges) become so high that the atoms effectively fall out of
the lattice. On the other hand, if the lattice becomes too deep the contrast decreases (see Fig.\
\ref{fig:snrfnlatdtcorr} below and also Fig.\ 5 in \cite{bpvenkatesh11}). A depth of $3
E_{\mathrm{R}}$ gives a reasonable compromise. Therefore, although in the rest of this paper we will
examine the effects of changing the various system parameters, we will always maintain the minimum
lattice depth at $3 E_{\mathrm{R}}$ (except in Fig.\ \ref{fig:snrfnlatdtcorr} and Fig.\ \ref{fig:linincdeplat}).
This also allows us to make comparisons between the effects of different parameter values upon,
e.g.\  the quantum fluctuations, whilst keeping the atomic meanfield dynamics as similar as
possible.

As the coupling $\beta$ is increased other effects appear apart from an increase in the contrast.
These effects are visible in Fig.\ \ref{fig:latticedepth&FT} (see also Fig.\ 2 in
\cite{bpvenkatesh09}). In Fig.\ \ref{fig:mftlatatsnrdip} we see that small-amplitude fast
oscillations of the lattice depth appear on top of the basic BO.  Referring to the Fourier transform of $s(t)$ plotted in Fig.\
\ref{fig:mftlatFTatsnrdip}, we see that the basic BO dynamics is governed by the fundamental
$\omega_{\mathrm{B}}$ and its low lying harmonics, whereas the fast oscillations are clustered
around the tenth harmonic (see inset) and include a continuum of frequencies with some peaks at half
harmonics. In this context it is important to bear in mind that the band gaps change continuously in
time as $q$ is swept through the Brillouin zone and so a range of frequencies is to be expected.

\begin{figure}
\includegraphics[width=0.9\columnwidth]{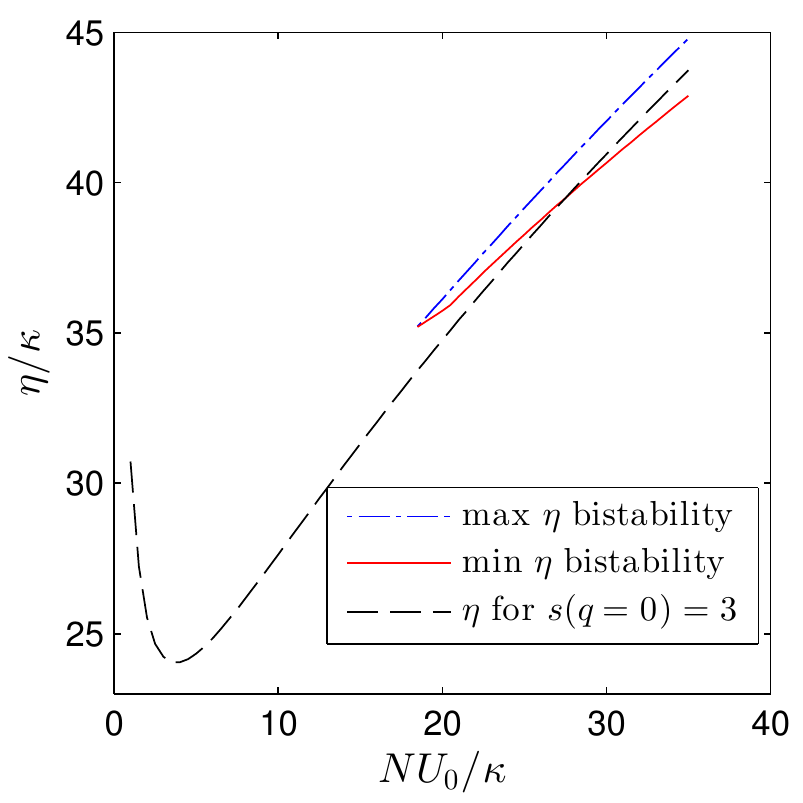}
\caption{(Color online) Plot of pump strength (dashed black line) required to maintain a minimum
lattice depth of $3
E_{\mathrm{R}}$ as a function of $NU_0/\kappa$.  The red (solid) and blue (dash dotted) lines
enclose the values of $\eta$ for which the
steady state photon number in the cavity is bistable for any value of the quasimomentum of the
atomic wave function. One sees that for
$NU_0/\kappa \sim 25$, the pump strength required to maintain the lattice depth leads to
bistability. Other parameters for the plot are
$\Delta_{\mathrm{c}} = -0.75 \, \kappa$, $\kappa =345 \, \omega_{\mathrm{R}}$, $N = 5 \times 10^4$}.
\label{fig:pumpstdc1_bistregime}
\end{figure}

The precision to which $\omega_{\mathrm{B}}$ can be measured in the  scheme proposed in this paper
depends upon the contrast. From the results shown in
\figref{fig:mftlatplotdc1_changecoup} it may therefore seem that in order to make the most sensitive
measurement possible one should choose $\beta$ to be as large as possible. However, this is false
for two reasons. One is the effect of quantum fluctuations due to measurement backaction which is
also controlled by $\beta$ and will be the focus of Section \ref{sec:snr}. Another reason, which
enters even at the meanfield level, is the possibility of bistability in cavity photon number for
large values of $\beta$ (when the pumping is sufficiently large). In \cite{brennecke08} this
bistability was studied experimentally in a uniform unaccelerated condensate, which in our language
has a quasimomentum $q=0$. In \cite{bpvenkatesh11} we studied this problem theoretically and
generalized it to include finite $q$: we showed that bistability arises from the appearance of
swallowtail loops in the bands. In the semiclassical picture of a BO the quasimomentum scans
adiabatically through the entire band and so when it encounters a swallowtail loop the system can
follow a branch that suddenly terminates at some later time, leading to fundamentally nonadiabatic
behavior \cite{bpvenkatesh09,wu00}. Hence, in a scheme to measure BOs, it would be better to be in a
parameter regime where the cavity is not bistable for any value of $q$. In
\figref{fig:pumpstdc1_bistregime} we plot the pump strength required to maintain the lattice depth
at a minimum value of $3 E_{\mathrm{R}}$  as a function of $\beta$. The red (solid) and blue
(dot-dashed) lines enclose the values of $\eta$ for which the steady state photon number in the
cavity displays bistability for at least some values of the quasimomentum. We see that for $\beta$
values as large as $25$ (at the fixed detuning $\Delta_{\mathrm{c}} = -0.75 \kappa$) one can avoid
bistability and get large contrast in the lattice depth evolution. 

Having emphasized that our choice for the pumping strength $\eta$ is guided by the tradeoff between
contrast and bistability according to Fig.\ \ref{fig:pumpstdc1_bistregime}, let us now explain how
we chose the rest of the system parameters used in the  calculations. There are three parameters we
hold constant throughout this paper; the first is the cavity damping rate $\kappa = 345 \,
\omega_{\mathrm{R}}$ which is the value realized in the experiment \cite{brennecke08}. As a guide to
the magnitude of the atomic recoil frequency  $\omega_{\mathrm{R}}$ used as the frequency unit, we
note that for $^{87}$Rb atoms in $780$nm light $\omega_{\mathrm{R}} = 2 \pi \times 3.8$ kHz. The
second constant parameter is the Bloch frequency $\omega_{\mathrm{B}} = \omega_{\mathrm{R}}/4$. The
gravitational force on $^{87}$Rb atoms in a $780 \mathrm{nm}$ lattice provides a Bloch frequency
very close to this value. Finally, unless specified otherwise, we keep the meanfield atom number
fixed at $N = 5 \times 10^4$ and vary $U_0$ in order to vary $\beta$. This last choice is motivated
by a scaling symmetry of the meanfield  equations [\eqnrange{eq:mftlight}{eq:mftatom}], which also
holds for the quantum operator equations [\eqnrange{eq:fluctphot}{eq:fluctatom}] below; solving the
coupled equations for the set of parameters $\{U_0,N,\eta,\Delta_{\mathrm{c}},\kappa\}$ is exactly
the same as solving them for $\{U_0r,N/r,\eta/\sqrt{r},\Delta_{\mathrm{c}},\kappa\}$, where $r$ is
some positive scaling factor. In both the scaled and unscaled versions the lattice depth $s(t) = U_0
\vert \alpha(t) \vert^2$ is maintained at the same value. Thus, each specific calculation one
performs represents a family of parameters. Our choice for $N$ keeps the atomic density dilute
enough in a typically sized cavity that the approximation of ignoring collisional atom-atom
interactions remains valid. There is some latitude in the choice of $\Delta_{\mathrm{c}}$, but the
contrast one obtains  at a given value of $\beta$ is larger for $\Delta_{\mathrm{c}}$ closer to the
cavity resonance. On the other hand, one also has to make sure that the effective cavity detuning 
\begin{align}
\dceff \equiv \Delta_{\mathrm{c}} - NU_0 \langle \cos^2(x) \rangle
\label{eq:coolingcondition}
\end{align}
is less than zero so that we are in the cavity cooling regime for the fluctuations \cite{szirmai09}
(see Section \ref{sec:excspect}). We set $\Delta_{\mathrm{c}}=-0.75 \kappa$ since we find that it
maximizes the contrast for the coupling value of $NU_0/\kappa = 1$. We will examine the effect of
changing the number of atoms $N$ and the minimum lattice depth when we examine the signal-to-noise
ratio in Section \ref{sec:snr}.

\section{Quantum Dynamics: theory}
\label{sec:fluctlin}

The approach we take to quantum dynamics is based upon a linearization about the meanfield solution,
retaining the quantum operators $\delta \hat{a}$ and $\delta \hat{\Psi}$ only to first order in the
equations of motion. This corresponds to the Bogoliubov level of approximation \cite{pethick,pita},
suitably generalized to describe coupled atomic and light fields.  A new feature of our problem in
comparison to previous linearization-based treatments of cavity-QED systems, e.g.\
\cite{horak01,szirmai10,steinke11}, is that our meanfield is time-dependent because of the BOs. This
means that the fluctuation modes, which must be orthogonal to the meanfield mode, also evolve in
time (not just their occupations).

Linearizing about the meanfield solution may appear to be an innocent strategy, but, as is well
known from the theory of Bose-Einstein condensation, care must be taken with such U(1) symmetry
breaking approaches  because they introduce a macroscopic (meanfield) wave function with a
particular global phase at the cost of particle number conservation \cite{lewenstein96}.  In
particular, when performing a linearization about the condensate there is always a trivial
fluctuation mode parallel to it with zero frequency (the ``zero mode'') which corresponds to
unphysical fluctuations of the global phase. These issues are even more acute when the condensate is
time-dependent and the boundary between condensate and fluctuation is further blurred 
\cite{castin97}.

The zero mode problem can be handled by only including fluctuations that are at all times orthogonal
to the meanfield.   We achieve this by applying the  projector $\hP(t)$ \cite{castin97,gardiner00}
\begin{equation}
\hP(t) =  \mathcal{I} - \vert \varphi(t) \rangle \langle \varphi(t) \vert \label{eq:proj} 
\end{equation}
so that 
\begin{align}
\delta \hat{\Psi}_{\perp}(x,t) & \equiv  \hat{P}(t) \, \delta \hat{\Psi}(x,t) \label{eq:perpfluct}\\
&= \int dy \left[ \delta(x-y) - \varphi(x,t) \varphi^{*}(y,t)\right] \delta \hat{\Psi}(y,t) \ .
\nonumber
\end{align} 
One consequence of this is that the commutator between atomic fluctuations is given by
\cite{fetter80} 
\begin{align}
 \left[ \delta \hat{\Psi}_{\perp}(x,t),\delta \hat{\Psi}_{\perp}^{\dagger}(y,t) \right] &= \langle x
\vert \hP(t) \vert y \rangle
\label{eq:atomcommute}\\
& = \delta(x-y) - \varphi(x,t) \varphi^{*}(y,t) \ . \nonumber
\end{align}
Unlike the usual bosonic commutator for the fluctuation field $\bthpsi$, this is time dependent.

Next, we transform the atomic fluctuation operator from the LF to the TF 
\begin{align}
 \bthpsi(x,t) = \delta \hat{\Psi}(x,t) e^{-iftx} 
\end{align}
which simplifies the calculation for the same reasons as mentioned in Section \ref{sec:meanfield} for
the meanfield. Since only ``vertical'' fluctuations between bands can occur, both the fluctuations
and the meanfield have the same quasimomentum (which in the TF is frozen at its initial value), and so we can expand the fluctuations and meanfield in the same basis
\begin{align}
\bvphi(x,t) &= \displaystyle \sum_{n} c_n(t) e^{i2nx} \label{eq:mftatomexp}\\ 
\bthpsip(x,t) &= \displaystyle \sum_{n} \dc_n(t) e^{i2nx} \ . \label{eq:atomopsplit}
\end{align}
This makes the numerics a little easier. Note that we have set the initial quasimomentum in these equations to $q_0=0$ without loss of generality. Meanwhile, back in
the LF, the quasimomentum  evolves according to the Bloch acceleration theorem given by
\eqnref{eq:blochaccthm}.

We can now write down the coupled equations of motion for the cavity and atomic fluctuation operators in the TF as
\begin{widetext}
\begin{align}
\refstepcounter{equation}
& i\frac{d }{dt}\delta \ha(t) = A(t) \delta \ha(t) + \sqrt{N} U_0 \alpha(t) \int
dx \cos^2(x) \left [\bvphi^{*}(x,t)  \bthpsip(x,t) + \bvphi(x,t)   \bthpsip^{\dagger}(x,t) \right] +
i \sqrt{2\kappa} \hxi(t)
\subeqn \label{eq:fluctphot}  \setcounter{subeqn}{1}  \\
& i \frac{\partial }{\partial t}\bthpsip(x,t) = \bmH(t)  \bthpsip(x,t) + \sqrt{N} U_0 \hP(t)
\cos^2(x) \bvphi(x,t) \left[\alpha^{*}(t)
\delta \ha(t) + \alpha(t) \dalp^{\dagger}(t) \right] \subeqn \label{eq:fluctatom}
\end{align} 
\end{widetext}
where $A(t) \equiv  -\Delta_{\mathrm{c}} + N U_0\langle \cos^2(x) \rangle(t) -i\kappa$.  The structure of these equations is such that without the Langevin term $\hat{\xi}(t)$ the operators $\delta \ha$ and $\bthpsip$ would be fixed at their initial values and so the  quantum parts of the fields would remain zero for all
time. The Langevin fluctuations appear as an inhomogeneous term in the cavity field equation and act
as a source that drives the evolution of $\delta \ha$ which in turn drives the evolution of $\bthpsip$ via
the atom-cavity coupling.

As pointed out in \cite{szirmai10}, the dynamics of the complex valued operators in the above
equations can be solved either by separating out their real and imaginary parts (optomechanics
approach) or by simultaneously solving the equations for the hermitian conjugates of the operators
(the Bogoliubov-de Gennes approach). We choose the latter. Collecting the fluctuations into the
column vector $\hat{R}(t) = \left( \dalp \ \dalp^{\dagger} \ \bthpsip \ \bthpsip^{\dagger} \right)^{T}$,
and the noise operators that act as source terms into the column vector $\hat{Z}(t) = \sqrt{2\kappa}
\left( \hxi \ \hxid \ 0 \ 0 \right)^{T}$, where $T$ denotes transposition, we obtain the operator matrix
equation
\begin{subequations}
\label{eq:fluctmatrixequation}
\begin{align}
i \frac{\partial}{\partial t} \hat{R} = \bM \hat{R} (t) + i \hat{Z}(t) \label{eq:mateqn}
\end{align}
with
\begin{widetext}
\begin{align}
\mathbf{M(t)} &= \begin{bmatrix} 
               A & 0 & \sqrt{N} U_0 \alpha V^{*} & \sqrt{N} U_0 \alpha V\\
		    0&-A^{*}&-\sqrt{N} U_0 \alpha^{*} V^{*} & -\sqrt{N} U_0 \alpha^{*} V\\
	      \sqrt{N} U_0 \alpha^{*} W(x) & \sqrt{N} U_0 \alpha W(x) & \hP \ \bmH(t) & 0\\
	      -\sqrt{N} U_0 \alpha^{*} W^{\dag}(x) & -\sqrt{N} U_0 \alpha W^{\dag}(x)&0&- \hP^{\dag}
\ \bmH(t)
              \end{bmatrix} \label{eq:fluctmat}
\end{align}
\end{widetext}
where we have introduced the operators 
\begin{align}              
V \cdot g(x) & \equiv \int dx \, \bvphi(x,t) \cos^2(x) g(x)  \label{eq:cosopfull} \\
W(x) & \equiv \hP(t) \cos^2(x) \bvphi(x,t)  \label{eq:Ycoupling}
\end{align}
\end{subequations}
i.e.\ $V$ is an integral operator that acts on a function $g(x)$. Since they fall on the
off-diagonals, the terms involving $V$ and  $W$ couple the cavity and atom fluctuations. Observe,
however, that in the linear approximation used here the atomic fluctuation operators $\bthpsip(x,t)$
are not directly coupled to the cavity fluctuation operators $\dalp(t)$ because this would lead to
terms which are of second order. Rather, the coupling between the two sets of quantum fields is
mediated by the meanfields $\alpha(t)$ and $\bvphi(x,t)$.

 The matrix $\bM(t)$ is non-normal, i.e.\ it does not commute with its Hermitian adjoint and its
left and right eigenvectors are not the same. However, it does have the following symmetry property:
a linear transformation $\mathcal{T}$ that swaps the first and second, and simultaneously, the third
and fourth rows, produces a matrix which is proportional to the complex conjugate of the original
\cite{szirmai09}
\begin{align}
 \mathcal{T}.\bM.\mathcal{T} = - \bM^{*}.
\end{align}
This symmetry, which is a general feature of Bogoliubov-de Gennes type equations \cite{wu03},
implies that the eigenvalues (and the associated eigenvectors) occur in pairs of the form $\pm
\omega_n + i \gamma_n$ i.e.\ with the same imaginary parts but with real parts of opposite sign. We
shall explore the spectrum of the fluctuation matrix $\mathbf{M}$ further in the Section
\ref{sec:excspect}. We also note that when written in matrix form the role of the projection
operator becomes clear since one can immediately see that the vectors $\left(0 \, 0 \, \bvphi(x,t) \, 0
\right)^{T}$ and $\left(0 \, 0 \, 0 \, \bvphi^{*}(x,t) \right)^{T}$ span the zero eigenvalue subspace of the
matrix $\bM$ and the trivial fluctuations live in this subspace.

The time evolution of the fluctuation operators is given by solving \eqnref{eq:mateqn}. However,
measurable observables are given by expectation values and correlation functions of these operators
rather than by the operators themselves. To this end we consider the covariance matrix  $\bC(t)$
associated with the vector $\hat{R}$
\begin{align}
 \bC_{jk}(t) \equiv \langle \hat{R}_j \hat{R}_k \rangle(t) \label{eq:covmatdef}.
\end{align}
Particular cases of $\bC_{jk}(t)$ (or more precisely, its sum) include the total number of photonic and atomic fluctuations
\begin{align}
\delta n(t) &= \langle \dalp^{\dagger}(t) \dalp(t) \rangle \label{eq:photfluctno} \\
 \delta N(t) &= \int dx \langle  \bar{\delta \hat{\Psi}}_{\perp}^{\dag}(x,t) \bthpsip(x,t) \rangle
\label{eq:atomfluctno}. 
\end{align}
The latter correspond to the number of atoms excited out of the meanfield component (i.e.\ the
atomic depletion).

To obtain the time evolution of the covariance matrix, consider the formal solution to
\eqnref{eq:fluctmatrixequation} \cite{liao11}
\begin{align}
\hat{R}(t) &= \bG(t,0) \hat{R}(0) + \bG(t,0) \int_0^t \bG^{-1}(\tau,0) \hZ(\tau) d \tau
\label{eq:fluctformsoln}
\end{align}
where $\bG(t)$ is a matrix satisfying:
\begin{align}
 \dot{\bG}(t,0) &= -i \bM(t) \bG(t,0) \ ;  \quad \bG(0,0) = \mathcal{I}  \, .
\label{eq:essentialmatevol}
\end{align}
We drop the dependence of $\bG$ on the initial time for notational convenience in what follows.
Inserting this formal solution in \eqnref{eq:covmatdef} we find
\begin{align}
\bC(t) &= \bG(t) \bC(0) \bG^{T}(t) + \bG(t) \Sigma(t) \bG^{T}(t) \label{eq:covmatsol}\\
\Sigma(t) & \equiv \int_0^{t} \int_0^{t} \bG^{-1}(\tau) \langle \hZ(\tau) \hZ(\tau^{\prime}) \rangle
[\bG^{-1}(\tau^{\prime})]^{T} d\tau
d\tau^{\prime}.
\end{align}
Using the property of the Langevin noise terms given in \eqnref{eq:whitenoise}, we can simplify
$\Sigma(t)$ as
\begin{subequations}
\begin{align}
\Sigma(t) & = \int_0^{t}  \bG^{-1}(\tau) D [\bG^{-1}(\tau)]^{T} d\tau\\
D_{jk}& \equiv 2 \kappa \delta_{j1}\delta_{k2}.
\end{align}
\end{subequations}
Our main numerical task is thus to solve the matrix differential equation given by
\eqnref{eq:essentialmatevol}. In addition, the matrix elements of $\bM(t)$ have to be computed from
the meanfields $\{\alpha(t),\bvphi(x,t) \}$ obtained by solving the coupled equations
\eqnref{eq:mftlight} and \eqnref{eq:mftatom}. These latter equations are simply a set of ordinary
differential equations that we solve using an adaptive time-step Runge-Kutta scheme. We then solve
the matrix differential equation for $\bG(t)$ using the same time grid as the meanfield solution.
For the matrix differential equation, and the associated solution for the covariance matrix
$\bC(t)$, we can again use a Runge-Kutta algorithm or exponentiate the fluctuation matrix $\bM(t)$
over the (small) time step intervals \cite{vanloan78}.

As a check on the results we can use the fact that the elements of the covariance matrix $\bC(t)$
have to obey the commutator relations \eqnref{eq:atomcommute} for the operators making up
$\hat{R}(t)$. For example, when the atomic operator is expanded as in \eqnref{eq:mftatomexp}, the
expectation value of the commutator relation \eqnref{eq:atomcommute} gives
\begin{align}
 \langle \dc_n \dc_m^{\dagger} \rangle - \langle \dc_m^{\dagger} \dc_n \rangle = \delta_{nm}
-\langle n \vert \bvphi(t)\rangle \langle
\bvphi(t) \vert m \rangle.  
\end{align}
The left hand side of this equation gives the difference between certain entries of the covariance
matrix and we can calculate its expected value (the right hand side) from the meanfield solution.
The degree of agreement between the two sides provides a measure of the accuracy of the fluctuation
calculation. In general, we find that the accuracy can be increased by taking smaller time steps.

In closing this section, we would like to point out that the meanfield solution already includes
Landau-Zener type tunnelling that causes the \emph{coherent} excitation of higher bands. By
contrast, the effect of Langevin fluctuations $\xi$ is to \emph{incoherently} populate different
bands. Within the linear approximation used here, the depletion of the atomic meanfield by quantum
excitations is not self-consistent, i.e.\ the meanfield is always normalized to $N$ atoms, whatever
the number of depleted atoms $\delta N$. The linearized equations are only valid
when $\delta N << N$, as expected from a Bogoliubov-type approach.

\begin{figure*}
\centering
 \subfloat[Real part of the quasiparticle spectrum as a function of quasimomentum 
]{\label{fig:spectrealplotdc1_coup2}\includegraphics[width=.45\textwidth]{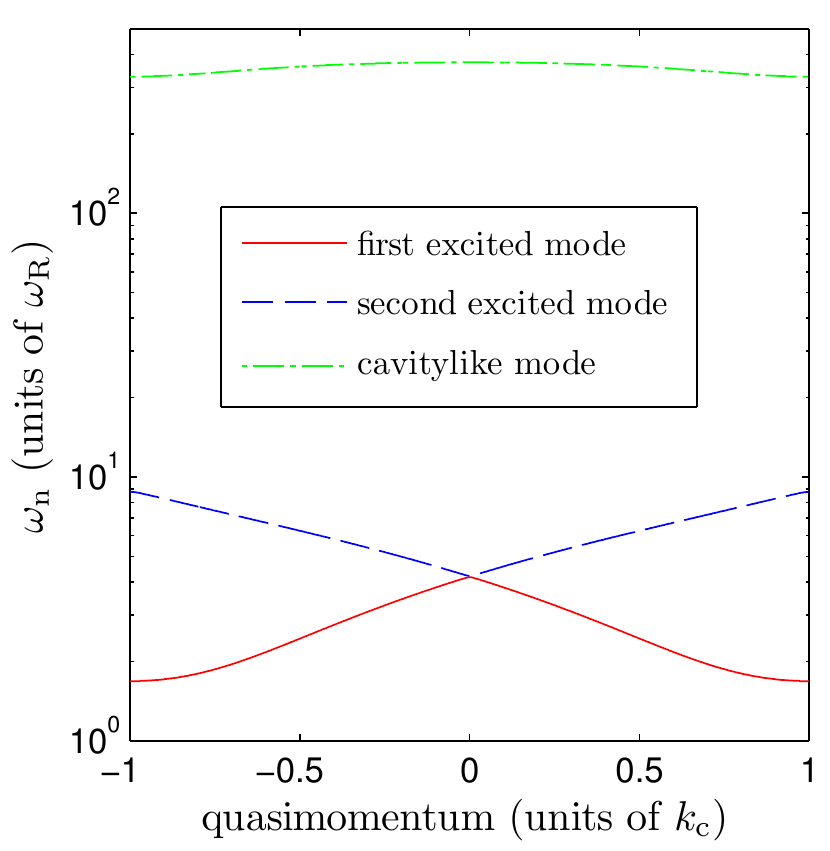}}
  \hfill
 \subfloat[Imaginary part of the quasiparticle spectrum as a function of
quasimomentum]{\label{fig:spectimagplotdc1_coup2}
\includegraphics[width=.45\textwidth]{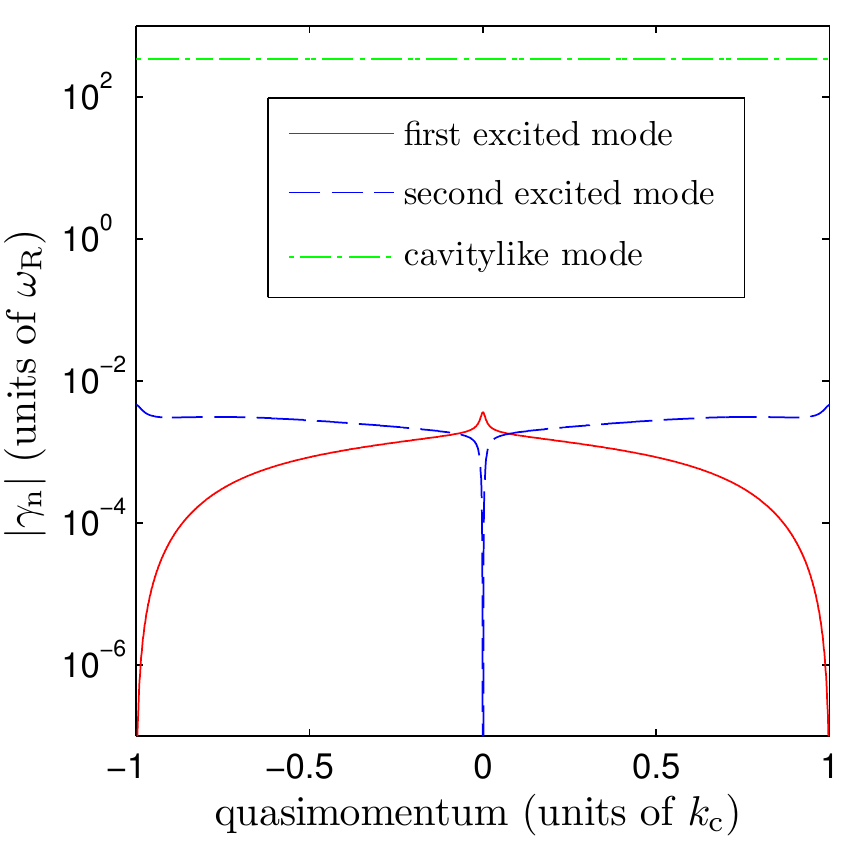}}
\caption{(Color online) Low lying levels in the quasiparticle spectrum (elementary excitations). The
frequency of the $n$th level is
generally complex  $\omega_{n}+i \gamma_{n}$. The parameters used in the plots are
$U_0 = 0.01 \, \omega_{\mathrm{R}}$, $\kappa =345 \, \omega_{\mathrm{R}}$, $\Delta_{\mathrm{c}} =
-0.75 \, \kappa$, and $N = 5 \times 10^4$.
The red (solid) and blue (dotted) lines correspond to hybridised
atom-cavity modes and generally have non-zero imaginary parts except at certain special points such
as at the band center and edges where
they can become marginally stable and decouple from the cavity. The green (dash dotted) line
corresponds to a cavity-like mode,  i.e.\ the
real part of its frequency is close to the effective detuning frequency $\dceff(q)$, and the
imaginary part is close to $-\kappa$.}
\label{fig:spectplots}
\end{figure*} 

\section{Spectrum of elementary excitations}
\label{sec:excspect}

Before presenting the results of the combined meanfield and quantum dynamics (see the next section),
we shall first examine the excitation spectrum of the atom-cavity system. The excitation spectrum
gives insight into the dynamics, heating effects, and will also be of use in explaining resonances that
affect the signal-to-noise ratio of the Bloch frequency measurement, a topic we will discuss in
Section \ref{sec:snrresults}.

We first note that there are two distinct types of excitation, and hence spectra. The coupled
atom-cavity band structure discussed in Section \ref{sec:meanfield} refers to meanfield excitations
which are labelled by a band index and a quasimomentum.  They involve every atom and photon
responding identically since, by the nature of the meanfield approximation, they are assumed to be
described by a single wave function $\varphi(x,t)$ and the coherent amplitude $\alpha(t)$,
respectively. On top of these, there are also elementary excitations or quasiparticles whose
energies are the complex eigenvalues $\pm \omega_{n}+i\gamma_{n}$ of the matrix $\mathbf{M}(t)$
given in Eq.\ (\ref{eq:fluctmat}). A clear description of the difference between the meanfield and
the quasiparticle spectra for a BEC in a (non-cavity) optical lattice can be found in
\cite{kramer03} and references therein. In the atom-cavity system the quasiparticles correspond to
single quanta of the combined fields and are thus \emph{polaritons}. The fact that they come in
pairs can be interpreted as an analogue of particles and antiparticles \cite{wu03}. 

Whereas the meanfield band structure is always real, the quasiparticle energies have an imaginary
part which comes from the leaking of the cavity field out of the cavity. If $\gamma_{n} <0$, we have
dynamical stability and $\vert \gamma_{n} \vert$ can be interpreted as the lifetime of the
quasiparticle. This damping effect has potentially very important applications in cavity-assisted
cooling \cite{horak01,gardiner01}. If, on the other hand, $\gamma_{n} >0$ we have dynamical
instability and heating.

 In general, the elementary excitations have a band structure all of their own, i.e.\ the solutions
of the Bogoliubov-de Gennes equations take the form of Bloch waves with a band index and
quasimomentum that can differ from that of the meanfield solution about which we are linearizing.
However, as discussed in Section \ref{sec:meanfield}, here we only allow excitations that preserve
the quasimomentum (vertical transitions), and thus our quasiparticles have the same quasimomentum as
their parent meanfield solution. Some examples of the quasiparticle band structure are plotted in
\figref{fig:spectplots} (see also Fig.\ \ref{fig:qpbandcrosscoup1} in Appendix \ref{app:longtimebehav}).

The eigenvectors of $\mathbf{M}$ can be classified into three kinds: cavity-like modes,
hybridised atom-cavity modes, and marginally stable modes \cite{szirmai09}. The cavity-like modes
(depicted by the green dash-dotted lines in \figref{fig:spectplots}) are close to being pure cavity
field modes with only a small atomic component. Hence, their eigenvalues have a real part with
magnitude close to the effective detuning $\dceff = \Delta_{\mathrm{c}} - NU_0 \langle \cos^2(x)
\rangle$, and an imaginary part approximately equal to $-\kappa$. The hybridized modes (depicted by
the red solid and blue dashed lines in \figref{fig:spectplots}) have some atomic and some cavity
field properties, whereas the marginally stable modes are purely atomic in nature with zero cavity
component. As we shall demonstrate below, the marginally stable modes occur at the points $q=0$ and
$q=\pm 1$, i.e. at the band center and edges, and their name derives from the fact that their
imaginary part is zero. 

The properties of the hybridised and marginally stable modes are determined by the sign of $\dceff$.
When $\dceff<0$ we find $\gamma_n<0$ and we are on the cooling side of the effective resonance. On
the contrary, when  $\dceff>0$ we find $\gamma_n>0$ and we are on the heating side. A calculation of
the dynamics on the heating side is not stable since the linearization will fail after a short time
due to the exponentially growing number of quasiparticles. Thus, the calculation of the spectra
serves a very useful purpose: it guides our choice of $\Delta_{\mathrm{c}}$ so as to ensure that we
are always on the cooling side of the resonance.

\begin{figure*}
\centering
  \subfloat[Excited atom fraction versus time for $NU_0/\kappa=0.1$]{\label{fig:fluctatapprox_smu0}
\includegraphics[width=.45\textwidth]{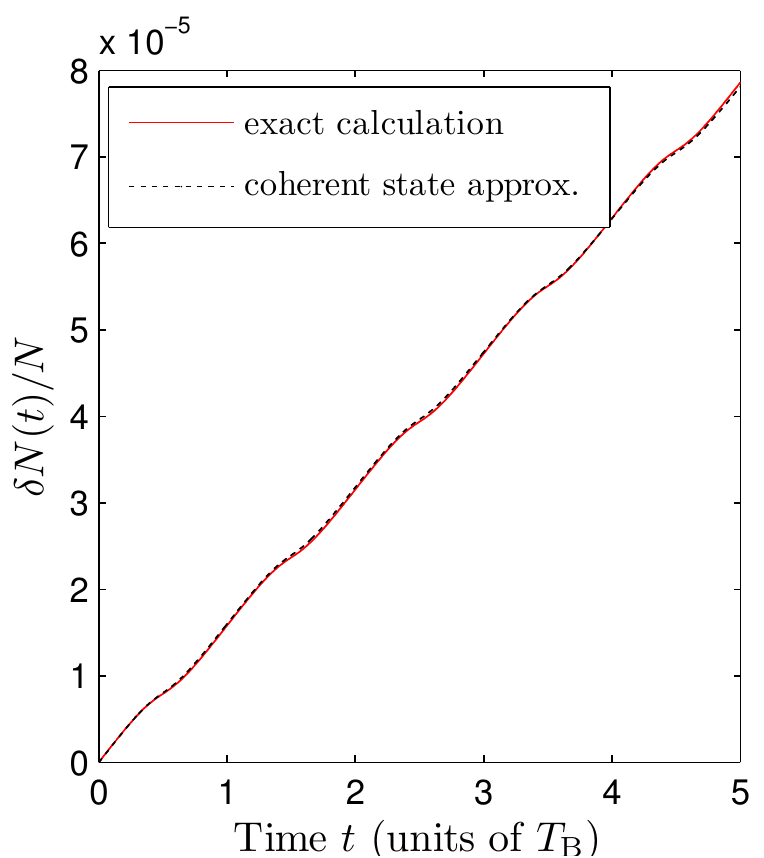}}
  \hfill
  \subfloat[Excited atom fraction versus time for $NU_0/\kappa=1$
]{\label{fig:fluctatapprox}\includegraphics[width=.45\textwidth]{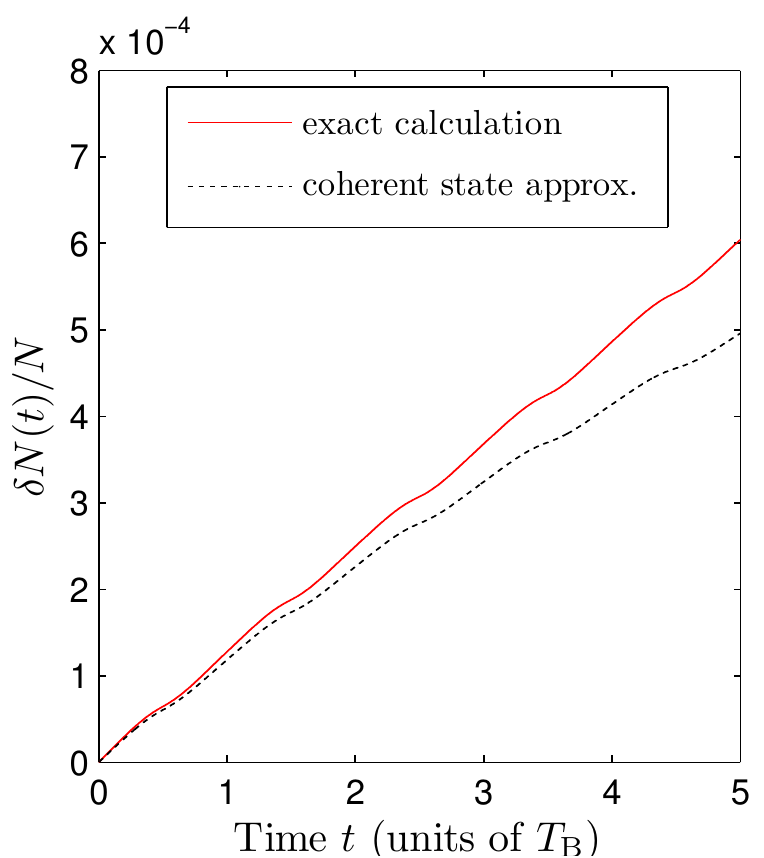}}
\caption{(Color online) Growth of the excited atom fraction over five BO periods for (a) weak and (b)
moderately strong atom-cavity
coupling. The red (solid) curves are given by solving the full quantum  problem in the form of
\eqnref{eq:essentialmatevol}, whereas the
black (dashed) curves are the result of treating the atomic modes as independent oscillators plus assuming that the quantum fluctuations in the light come purely from vacuum shot noise, i.e.\ the coherent state approximation. The meanfield  dynamics for (b) is given by the red (solid) curve in \figref{fig:mftlatplotdc1_changecoup}. The atom heating rate in these figures oscillates because it is lower at the Brillouin zone edges than at the center. Referring to \figref{fig:spectplots} we see that at the zone edges the quasiparticle mode with the smallest real part (red solid curve) becomes marginally stable, i.e.\ the cavity light field part and the atomic part decouple.}
\label{fig:dNplots}
\end{figure*} 

In \figref{fig:spectrealplotdc1_coup2} the red (solid) and blue (dashed) lines give the magnitudes
of the real parts of the frequencies of the two lowest quasiparticle eigenmodes as a function of quasimomentum.
The magnitudes of the imaginary parts are plotted in \figref{fig:spectimagplotdc1_coup2}. Notice
that the imaginary part of one of the modes goes to zero at the band edges (red solid line) and the
other goes to zero at the band center (blue dashed line). This implies that these excitations are
only marginally stable at those specific values of the quasimomentum. The vanishing of the imaginary
part of the frequencies at these points can be understood as follows: a Bloch wave with $q=0$
($q=1$) is even (odd) about the center of a single cell $[0,\pi]$ of the $\cos^{2}(x)$ potential.
Since the atomic meanfield solution $\bar{\varphi}(x)$ is a Bloch wave it has well defined parity at
these points. The same is also true for the atomic part of quasiparticle eigenmodes $\bthpsip(x)$ [and of course $\bthpsip^{\dag}(x)$] of the fluctuation matrix $\bM$, for these are also Bloch waves. In this case the integral $\int dx \bar{\varphi}^{*}(x)
\cos^2(x)  \bthpsip(x)$ will sometimes vanish identically because the integrand can contain
functions with opposite parity. Examination of the fluctuation matrix $\mathbf{M}$ given in Eq.\
(\ref{eq:fluctmat}) shows that it is exactly this integral that controls the weight of the cavity
part of the quasiparticle eigenmodes, and so at $q=0,\pm1$ we can have undamped quasiparticles with
$\gamma = 0$. For other values of quasimomentum the meanfield wave function has no particular
parity and there are no marginal modes.

\section{Quantum Dynamics: results} 
\label{sec:fluctdyn}

In this section we present results from the numerical solution of the quantum equations of motion.
We assume that at $t=0$ the fluctuation fields corresponding to $\dalp$ and $\bthpsip$ are in their vacuum states and
expand the atomic part in the basis given in \eqnref{eq:atomopsplit}. This allows us to construct
the covariance matrix \eqnref{eq:covmatdef} $\bC(t=0)$ which we then evolve to later times using
\eqnref{eq:essentialmatevol}. In order to perform this task we need the fluctuation matrix $\bM(t)$
as a function of time which in turn requires the meanfield solution $\{ \bvphi(x,t),\alpha(t)\}$ as
input. We therefore solve the meanfield dynamics on the same discretized time grid in parallel
with the computation of \eqnref{eq:essentialmatevol}. 

\begin{figure*}
\centering
  \subfloat[Excited atom fraction versus time for different values of $\beta$.]
{\label{fig:fluctphatnuma}\includegraphics[width=.45\textwidth]{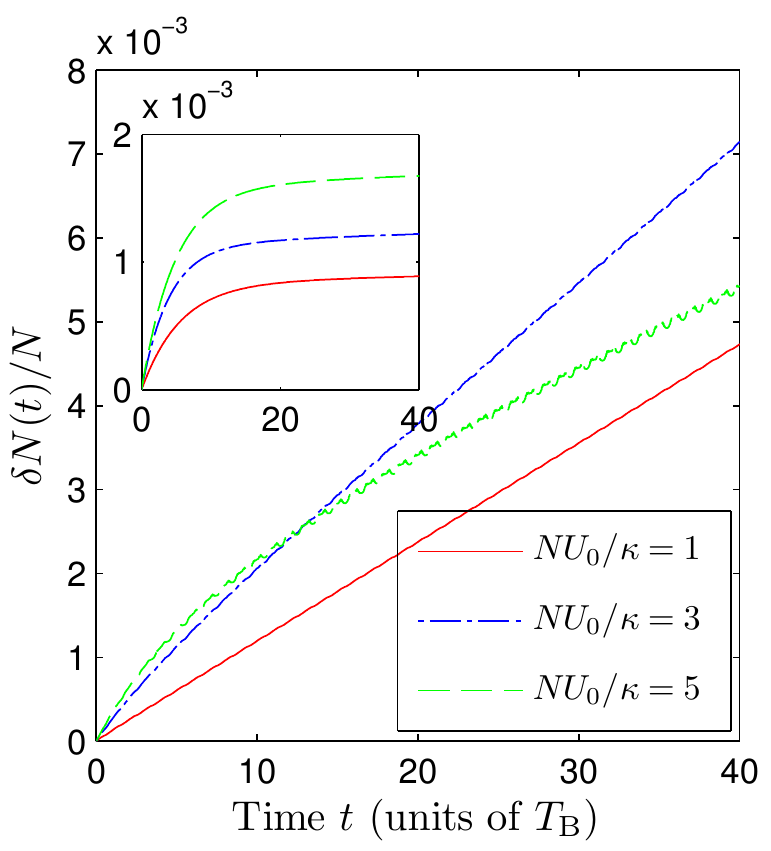}}
  \hfill
 \subfloat[Excited photon number versus time for different values of $\beta$.]
{\label{fig:fluctphatnumb} \includegraphics[width=.45\textwidth]{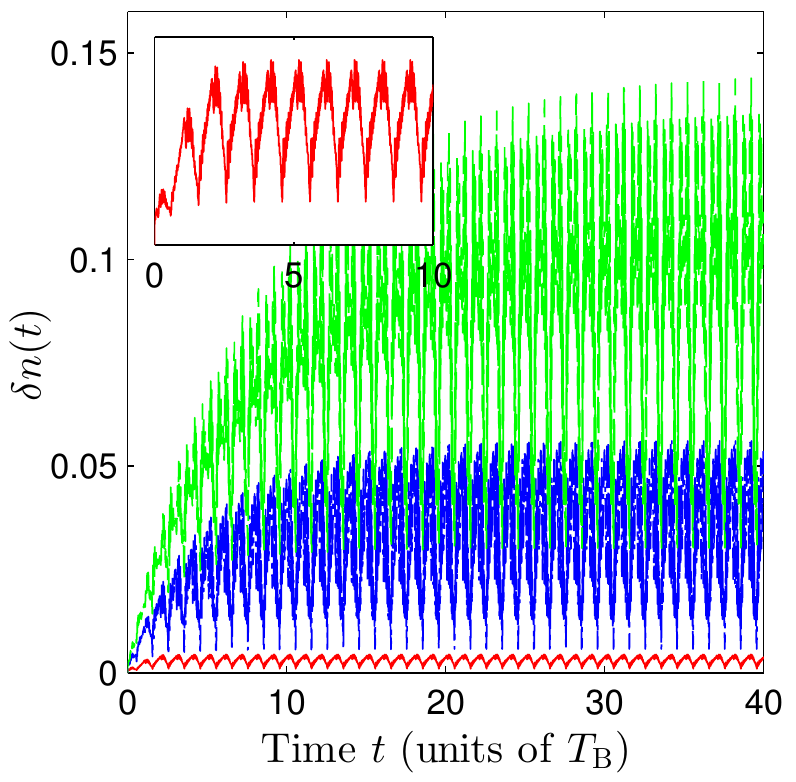}}
\caption{(Color online) Plots of (a) atomic  and (b) photonic fluctuation occupation number over
40 BOs calculated using a numerical solution of \eqnref{eq:essentialmatevol}. The inset in (a) shows $\delta N/N$ as a function of time for the case \emph{without} an external force, and hence without BOs, whereas the main body of (a) shows the results with an external force: $\delta N/N$ quickly reaches a steady state in the former but not in the latter case. In
(b) the lowest curve (red) is for $\beta=1$, the middle curve (blue) is for $\beta=3$ and the highest curve (red)
is for $\beta=5$. The inset in (b) shows a close up of the photonic fluctuation number as a
function of time for $\beta=1$. It can be seen how after a transient period the photonic fluctuation
number oscillates at the Bloch period, thereby mirroring the meanfield dynamics.}
\label{fig:dNplots2}
\end{figure*}

Once we have computed $\bC(t)$, we can use it to calculate
the physical quantities of interest, such as the number of atomic excitations $\delta N(t)$, as defined in
\eqnref{eq:atomfluctno}. This should not be confused with the number of quasiparticles, which are generally made up of both atomic and cavity field components. If it were not for the
Langevin noise, the evolution would be perfectly coherent and $\delta N$ would be zero. However, the
presence of Langevin noise in the electromagnetic field generates atomic excitations via the
atom-cavity coupling.   In \figref{fig:dNplots} we plot the fraction $\delta N (t)/N$  as a function
of time for five Bloch periods for two different coupling values. The red (solid) curves are given
by a full solution of the quantum equations, whereas the black (dashed) curves are made with a
coherent state approximation for the cavity field, which will be outlined below and is discussed in
more detail in Appendix \ref{app:shotnoise}. The gradient of the curves in Figs.\ \ref{fig:dNplots} gives the heating rate and we note from \figref{fig:fluctatapprox} that the coherent state approximation slightly underestimates the true heating rate for atoms. The behaviour of $\delta N (t)/N$ over longer times (40 Bloch periods) is shown in Fig.\ \ref{fig:fluctphatnuma} and the equivalent quantity for the photons is shown in Fig.\ \ref{fig:fluctphatnumb}. We see that while the number of photons excited out of the meanfield saturates due to the damping by photon loss from the cavity, the atoms maintain a finite heating rate over all times we have investigated. This is perhaps surprising because we are on the cooling side of the resonance [see Eq.\ (\ref{eq:coolingcondition})] all the time (despite the modulations in the effective cavity detuning due to the BOs). In the inset in Fig.\ \ref{fig:fluctphatnuma} we show the case without BOs, and as can be seen, we recover the cooling. The presence of BOs clearly counteracts the cooling to some degree and prevents $\delta N (t)/N$ from reaching a steady state. This residual heating effect is analyzed in detail in Appendix \ref{app:longtimebehav}, but it turns out to be due to the transport of quasiparticles to higher energy states by Landau-Zener transitions that are driven by the BOs. A finite heating rate implies that at long propagation times the validity of our linearized approach will break down because $\delta N /N$ will no longer be small. However, for all the times and coupling strengths considered in this paper we have $\delta N /N < 1/100$.

In order to gain further insight into the dynamics, let us develop a semi-analytic model that we can compare against the exact results: it will allow us to see when atom-light correlations are important. The model makes two approximations: firstly we treat each eigenstate of the instantaneous meanfield Hamiltonian Eq.\ (\ref{eq:spHtrans}) as an independent oscillator mode uncoupled from the other modes, and secondly we approximate the state of the light inside the cavity as a coherent state. Coherent states have a noise spectrum that corresponds to the vacuum and so neglect correlations with the atoms. In fact, the second approximation follows naturally from the first as we show in Appendix \ref{app:shotnoise}.  The results of the approximate model are the black dashed curves in \figref{fig:dNplots}. The agreement with the exact results at weak coupling ($\beta=0.1$) is excellent, but begins to break down over time at stronger coupling ($\beta=1$), thereby revealing the dynamic generation of correlations. In fact, as mentioned in the Introduction, the heating rate of a cloud of cold atoms inside a cavity has been measured by Murch \textit{et al} \cite{murch08},  and they found it to be consistent with the predictions of vacuum noise. The new feature in our problem is that the effective cavity drive detuning $\dceff(t)$, which appears in the phase terms in Eq.\ (\ref{eq:independentshotnoise}), is Bloch periodic due to the meanfield dynamics.  

To motivate the coherent state approximation consider the exact solution to the  first order inhomogeneous differential equation for cavity field fluctuations Eq.\ (\ref{eq:fluctphot}), which can be formally written as 
\begin{align}
 \dalp(t) &=  e^{-i \int_{0}^{t} dt' A(t')} \times  \label{eq:lightformalsol} \\
    \int_0^t dt' & e^{i \int_{0}^{t'} dt'' A(t'')} [\sqrt{2 \kappa} \hat{\xi}(t') -i\sqrt{N} U_0
\alpha(t') \hat{X}(t')]   \nonumber 
\end{align}
where $\hat{X}(t) \equiv \int dx \, \bvphi^{*}(x,t) \cos^2(x) \bthpsip(t) + \mathrm{h.c.}$, and
$A(t) = -\dceff(t) - i \kappa$, as above.
We see that the cavity field fluctuation has two distinct contributions: the first term depends on the Langevin noise which accounts for vacuum fluctuations, whilst the second term depends on the state of the atoms. The coherent state approximation consists
of dropping the latter term in favour of the former to give
\begin{align}
\dalp(t) \approx \sqrt{2 \kappa} \int_{0}^{t} dt' e^{(i\dceff(t)-\kappa)(t-t') } \hat{\xi}(t') \ .
\label{eq:independentshotnoise}
\end{align}
In writing $\dalp(t)$ in this way we have taken advantage of the fact that the cavity decay rate
$\kappa$ is much faster than the  frequency $\omega_{\mathrm{B}}$ at which $\dceff(t)$ evolves, and
so the integrand is appreciable only for times  $t-t' \lesssim \kappa^{-1}$ during which $\dceff$ is
a constant and can be evaluated at time $t$. The regime of validity of the coherent state approximation can be estimated from its derivation which requires $r \equiv \sqrt{N}U_0 \vert \alpha(t) \vert/\sqrt{2 \kappa}<<1$. Note that $r^2 =
\beta s(t)/2$. In our earlier discussion (Section \ref{sec:mftresults}) of desirable parameters, we
stipulated a minimum lattice depth of $s(t) \sim 3 \, \omega_{\mathrm{R}}$, which implies that the
validity of the coherent state approximation here is contingent upon $\beta \ll 1$, i.e.\ this is a
weak coupling approximation.

The assumption of uncorrelated vacuum noise is a standard one in the field of cavity optomechanics  \cite{marquardt07,braginsky77,marquardt08,clerk10,clerk10b}. The paradigmatic example  is a cavity with one end mirror attached to a spring or cantilever i.e.\ a harmonic oscillator driven by radiation pressure. Although ultracold atoms in a very shallow lattice in a cavity can be mapped onto this system \cite{murch08,brennecke08,szirmai10,purdy10,nagy09}, that is not the case here because the atomic Bloch states do not map faithfully onto a single harmonic oscillator. Nonetheless, we have obtained our approximate model by applying a similar philosophy by mapping onto a collection of \emph{independent} oscillators (the eigenstates of $\mathcal{H}$). The coherent state approximation for the atomic excitation occupation $\delta N(t)$ that is
plotted as the black (dashed) curves in \figref{fig:dNplots} is the sum over the occupation
numbers of these independent oscillator modes $\delta N(t) =  \sum_j \delta N_j(t)$.  
The details of the mapping are presented in the Appendix \ref{app:shotnoise} and here we shall only sketch out the main idea
which is to consider the noise as a perturbation to the oscillator
dynamics, and then use Fermi's golden rule to calculate the noise induced transition rates amongst each oscillator's states.
This leads to a rate equation describing the occupation
number dynamics for each oscillator \cite{marquardt08}
\begin{align}
 \frac{d \langle \delta N_j \rangle}{dt} = \left( \Gamma_{uj} - \Gamma_{dj} \right) \langle \delta
N_j \rangle + \Gamma_{uj} \ ,
\label{eq:indoscexceqntext}
\end{align}
which is Eq.\ (\ref{eq:indoscexceqn}) in Appendix  \ref{app:shotnoise}. In this expression
$\Gamma_{uj}$ and $\Gamma_{dj}$ are the transition rates ``up'' and ``down'' for the $j$th
oscillator and they are proportional to $S_{\mathcal{FF}}(-\omega_j)$ and
$S_{\mathcal{FF}}(\omega_j)$, respectively, where $S_{\mathcal{FF}}(\omega)$ is the spectral density
of force fluctuations (shot noise power spectrum). Thus, each oscillator is driven and damped by
vacuum noise, with the rates of driving and damping being time dependent (due to the meanfield BO
dynamics).

In the next two sections we examine the effects of the fluctuations upon a precision
measurement, i.e.\ how the fluctuations put a limit on how large a value of $\beta$ can be chosen
for a precision measurement.

\section{Signal-to-Noise Ratio: theory} 
\label{sec:snr}

We now explore how the inclusion of quantum noise affects the precision measurement proposal in
\cite{bpvenkatesh09}. Recall the basic idea shown schematically in  \figref{fig:schemepic}: a cloud
of cold atoms undergoes Bloch oscillations (e.g.\ due to gravity) inside a Fabry-Perot cavity, and
the light field transmitted through the cavity is measured in order to determine the Bloch
frequency. In order to quantify the measurement performance we will compute the signal-to-noise
ratio using standard input-output theory \cite{walls08}. 

Let us consider a double sided cavity with mirrors with matched reflectivities providing equal
amplitude damping rates of $\kappa/2$.  The quantum part of the input fields for both the top
(driving side) and bottom (detection side) mirrors is given by the electromagnetic vacuum. Since we
are not going to consider classical fluctuations of the driving laser we do not include a classical laser field contribution in the input field , but introduce it via the hamiltonian in
\eqnref{eq:hamiltonian}. In our consideration of system dynamics in earlier sections we implicitly
assumed a single sided cavity giving an amplitude damping rate of $\kappa$, and associated with this
decay is a Langevin noise term $\sqrt{2 \kappa} \hxi(t)$. In a double sided cavity we have two
\emph{independent} noise terms of the form $\sqrt{\kappa} \{ \hxi_{t}(t),\hxi_{b}(t) \}$. It can be
shown that the dynamics of the intracavity system (both meanfield and fluctuations) are independent
of whether we assume a double sided or single sided cavity as long as we divide the net damping
equally among the two mirrors (provided they have matched reflectivities). The transmitted light
field is the output field at the bottom mirror which is related to the input field at the bottom
mirror as
\begin{align}
\aout(t) = -\ain(t) + \sqrt{\kappa} \ha(t) = -\hxi_b(t) + \sqrt{\kappa} \ha(t)
\end{align}
where $\ain$ and $\aout$ in this equation refer to the fields at the bottom mirror. The transmitted
photon current is given by the operator $\iout(t) = \aoutd(t) \aout(t)$, where again $\aout$ refers
to the field leaving the bottom mirror.  

An experimentally straightforward method for measuring the Bloch frequency consists of recording the
transmitted photon current using a photodetector. It is useful to consider the Fourier transform of
the data \cite{peden10}
\begin{align}
 \hat{N} (\omega,T) = \int_0^T dt \cos (\omega t) \iout(t)  \label{eq:signalop}
\end{align}
and define the signal-to-noise ratio for the measurement as  
\begin{align}
\mathrm{SNR}  \equiv \frac{\vert \langle \hat{N} (\omega,T) \rangle\vert^2}{\Delta N^2(\omega,T)}
\label{eq:SNRexp}  
\end{align}
where $\Delta N^2(\omega,T) \equiv \langle(\hat{N}-\langle \hat{N} \rangle )^2 \rangle$. Thus, the
SNR is the ratio of the spectral density of the photon current to its variance and provides one
measure of the sensitivity of the scheme. 

Let us first evaluate the SNR for a classical cavity field  $\ha(t) = \alpha(t)$. In this case one
finds that the signal amplitude and
variance are given by
\begin{align}
 \langle \hat{N}(\omega,T)  \rangle & = \kappa \int_0^T dt \cos(\omega t) \vert \alpha(t) \vert^2
\label{eq:signalmft}\\
\Delta N^2(\omega,T)  &= \kappa \int_0^T dt \cos^2(\omega t) \vert \alpha(t) \vert^2
\label{eq:varmft} \, .
\end{align}
In order to obtain an approximate magnitude for the SNR we further assume that the detection rate
goes as $ \approx R (1+ \epsilon \cos[\omega_{\mathrm{B}} t])$ \cite{bpvenkatesh09}, where
$\epsilon$ is the contrast parameter defined in \eqnref{eq:contrastdef}. Setting the classical
photon current $\kappa \vert \alpha(t) \vert^2$ in the above formulae equal to this detection rate 
gives
\begin{align}
\mathrm{SNR}(\omega_{\mathrm{B}},T) \approx \frac{\epsilon^{2} R T}{2} \ .  \label{eq:snrshotnoise}
\end{align}
Despite appearances, this result \emph{does} include quantum noise to a certain degree because
without the Langevin operators the variance given in Eq.\ (\ref{eq:varmft}) would have been zero,
i.e.\ even when the cavity field is classical the output field contains a quantum part $\aout =
\sqrt{\kappa}\alpha(t) - \hxi_b(t)$. Thus, the above calculation includes detector shot noise, also known as
measurement imprecision \cite{clerk10}, but neglects the effect of quantum fluctuations on the
coupled dynamics inside the cavity, i.e.\ quantum measurement backaction. Note that this is a
different approximation from the coherent state approximation used in Section \ref{sec:fluctdyn},
where quantum fluctuations were included in the cavity dynamics by using a Glauber coherent state,
i.e.\ a state with vacuum noise, for the cavity field, albeit one whose fluctuations are unaffected by
the presence of the atoms.

\begin{figure*}
\centering
  \subfloat[Signal-to-noise ratio as a function of $\beta= NU_0/\kappa$.
]{\label{fig:snrfnofcoup1}\includegraphics[width=.45\textwidth]{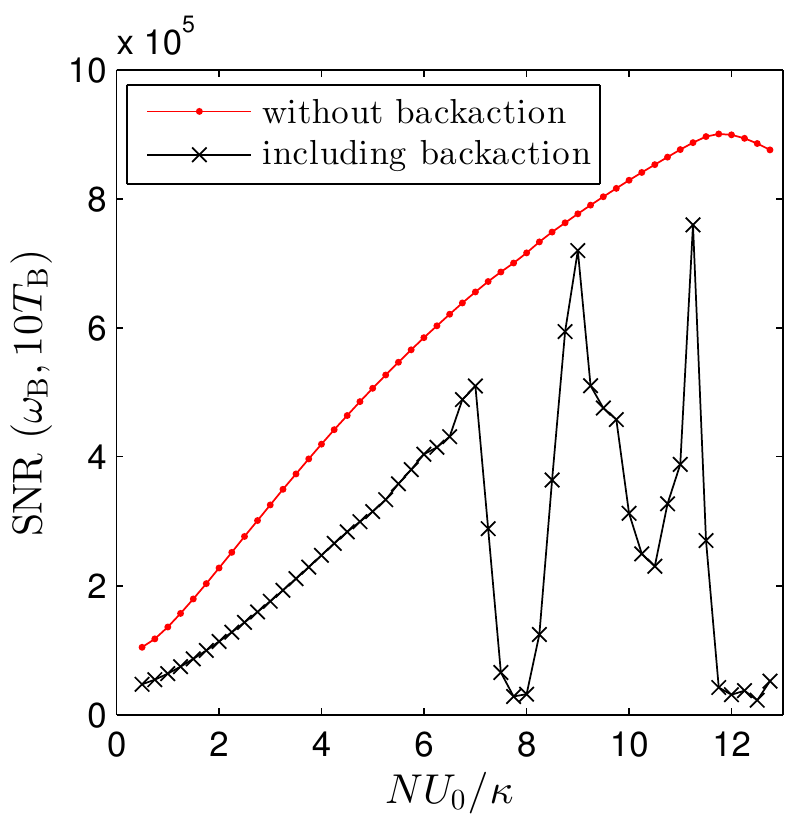}}
  \hfill
 \subfloat[Signal-to-noise ratio as a function of integration
time]{\label{fig:snrfnoftimenearpeak}
\includegraphics[width=.45\textwidth]{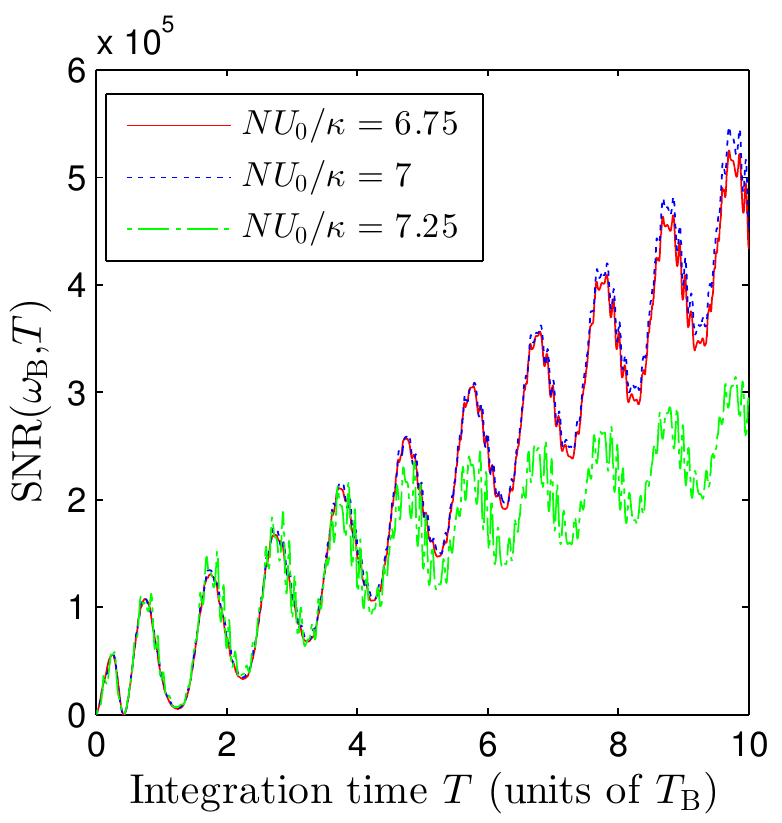}}
\caption{(Color online) Plots of the SNR as a function of (a) coupling strength $\beta$, and (b) integration time $T$ for different
values of $\beta$. In (a) the SNR was computed for an integration time of 10 Bloch periods
($T_B$) and the red (dots) curve gives the
meanfield dynamics plus detector shot noise result, whilst the black (crosses) curve includes
measurement backaction, i.e. the effect of
quantum fluctuations upon the coupled atom-cavity dynamics. In (b) the red (solid) and blue (dotted)
curves lie almost on top of each other
and correspond to values of $\beta$ just before the first dip in the SNR shown in
\figref{fig:snrfnofcoup1}, whereas the green (dash-dotted)
curve corresponds to a value of $\beta$ in the dip. For all plots the minimum lattice depth was 
3$E_\mathrm{R}$. Other parameters are the
typical ones mentioned in the text.}
\label{fig:snrplots1}
\end{figure*} 

\begin{figure*}
\centering
  \subfloat[Signal-to-noise ratio as a function of min.\ lattice depth
]{\label{fig:snrfnlatdtcorr}\includegraphics[width=.45\textwidth]{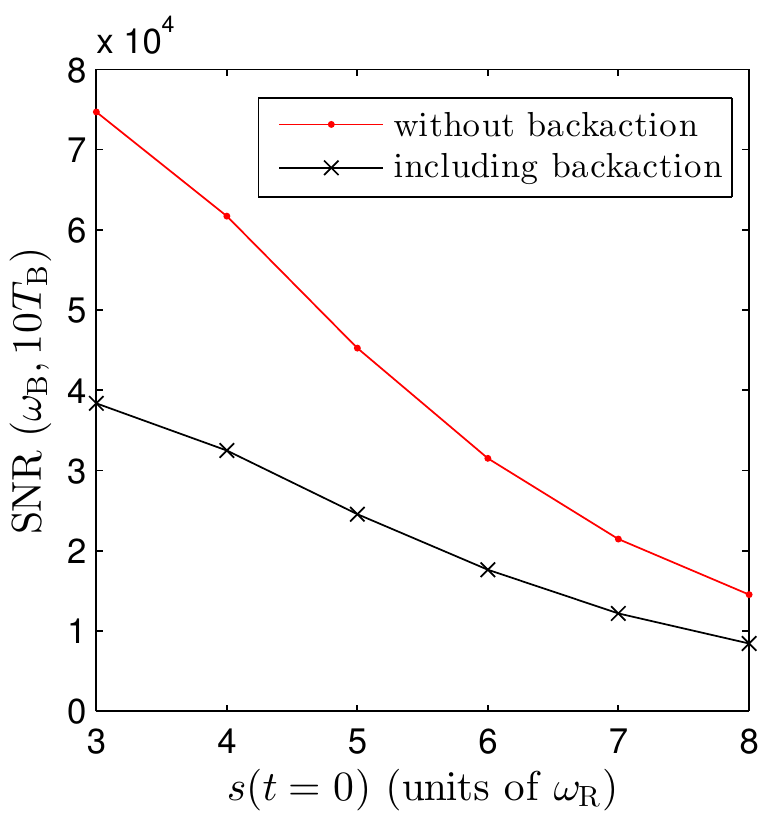}}
  \hfill
 \subfloat[Signal-to-noise ratio as a function of atom number]{\label{fig:snrfnNdtcorr}
\includegraphics[width=.45\textwidth]{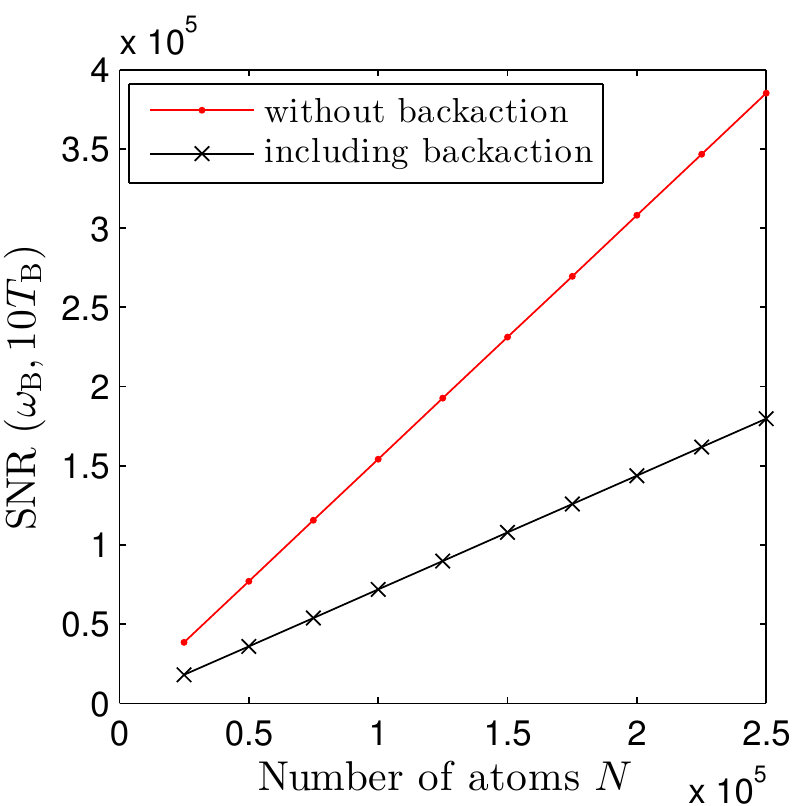}}
\caption{(Color online) Plots of the SNR as a function of (a) minimum lattice depth $s(t=0)$, and (b) atom number $N$. In
both plots the red (dots) curves were computed from meanfield theory plus detector shot noise, and
the black (crosses) curves were computed
including quantum measurement backaction. For all points $NU_0/\kappa=1$ and the signal is
integrated over 10 Bloch periods. In (a) it is
evident that the SNR decreases in both cases for larger lattice depths. In (b) it is evident that
the SNR increases linearly as a function
of $N$ in both cases.}
\label{fig:snrplots2}
\end{figure*} 

The SNR given by Eq.\ (\ref{eq:snrshotnoise}) predicts that the sensitivity of the scheme can be
increased indefinitely by increasing the mean total number of photons collected $RT$ and also the
contrast $\epsilon$. The former effect is the standard one expected from the general theory of
measurements with uncorrelated fluctuations. The latter is intuitively plausible too, but, however,
can not be the whole truth because, as stated above, it neglects the effect of measurement
backaction upon the dynamics which is expected to become important at larger values of $\beta$. 
When fluctuations are included $\ha(t) = \alpha(t) + \dalp(t)$,  and the mean signal amplitude is
given by \begin{align} \langle \hat{N}(\omega,T) \rangle = \kappa \int_0^T dt \cos(\omega t)
\left(\vert \alpha(t) \vert^2 + \langle \dalpd(t) \dalp(t) \rangle \right) \, .
\label{eq:signalfull} \end{align} In fact, this is not so very different from the meanfield photon
number given by Eq.\ (\ref{eq:signalmft}) because we are by design working in a regime where the
meanfield dominates the fluctuations. However, the same is not true of the signal variance. The
expression for the signal variance including fluctuations is cumbersome and is presented  in
\eqnref{eq:varoutphotds} in Appendix \ref{app:corrcalc}. For present purposes it is enough to note
that it includes  a collection of terms that depend on integrals over two-time correlations of the
photon fluctuations. These two-time correlations are challenging to evaluate numerically not only
because the fluctuations occur on time scales $\kappa^{-1}$ much shorter than the BOs, but also
because they require the storage and manipulation of data at two times. Furthermore, the continuous
driving by the BOs means that the correlations are not stationary in time, i.e.\ they do not just
depend on $t_{1}-t_{2}$, and this forces us to calculate the SNR in parallel to the system dynamics
starting at $t=0$. Unfortunately, due to limited computing power, we have only been able to track
the SNR over ten Bloch periods which is certainly shorter than the coherence time of the BOs for the
parameters we use. An actual experiment would, of course, not suffer from this limitation and would
benefit from running until the BO coherence time is reached.  The main steps of our algorithm for
calculating the two time correlations are provided in Appendix \ref{app:corrcalc}.

\section{Signal-to-Noise Ratio: results} 
\label{sec:snrresults}

We now show how the SNR depends on the various system parameters. Due to the size of parameter
space, this will not be an exhaustive study, but rather an \emph{ad hoc} choice that nevertheless we
hope is experimentally relevant.   We begin by looking at the SNR as a function of the coupling
parameter $\beta = NU_0/\kappa$. In \figref{fig:snrfnofcoup1} we plot the SNR evaluated at
$\omega_{\mathrm{B}}$ for an integration time of 10 Bloch periods. We change $\beta$ by increasing
$U_0$ but also change $\eta$ to maintain the same minimum lattice depth of $3 E_{\mathrm{R}}$
throughout. The results without measurement backaction (i.e. the dynamics in the cavity is purely
meanfield) are plotted by the red (dots) curve which monotonically increases until about $\beta=12$.
The initial increase of the SNR with $\beta$ is in line with expectations based on
\eqnref{eq:snrshotnoise}. The turnover of the red curve near $\beta=12$ is in a sense an artefact
that arises from having evaluated our SNR at $\omega_{\mathrm{B}}$: it so turns out from the
meanfield solution that for $\beta>12$ the fraction of the power in the fundamental of $s(\omega)$
begins to decline and is diverted to higher harmonics. However, there is no real reason other than
simplicity to only consider SNR$(\omega_{\mathrm{B}})$ (any harmonic of $\omega_{\mathrm{B}}$ gives
information about the applied force and inclusion of all of them in the data analysis would extract
the maximum possible information from the measurement). The full calculation including measurement
backaction is plotted by the black (crosses) curve. The first thing to notice is that measurement
backaction always lowers the SNR. Secondly, the full SNR monotonically increases only until $\beta
\approx 7$, and thereafter suffers from dramatic dips which we explain below as being due to
resonances with quasiparticle excitation energies.  These two observations are the main results of
this paper. In \figref{fig:snrfnoftimenearpeak} we plot the SNR as function of the total integration
time $T$ for three values of $\beta$, two before the first dip in the SNR and one in it. This plot
further illustrates that for $\beta > 7$ there is a dramatic lowering of the SNR. The BO dynamics
are also clearly visible due to the fact that the contrast is periodically growing and shrinking as
the lattice depth grows and shrinks.

\begin{figure*}
\centering
  \subfloat[Lowest excitation frequency for three different values of $\beta$
]{\label{fig:specreseff}\includegraphics[width=.45\textwidth]{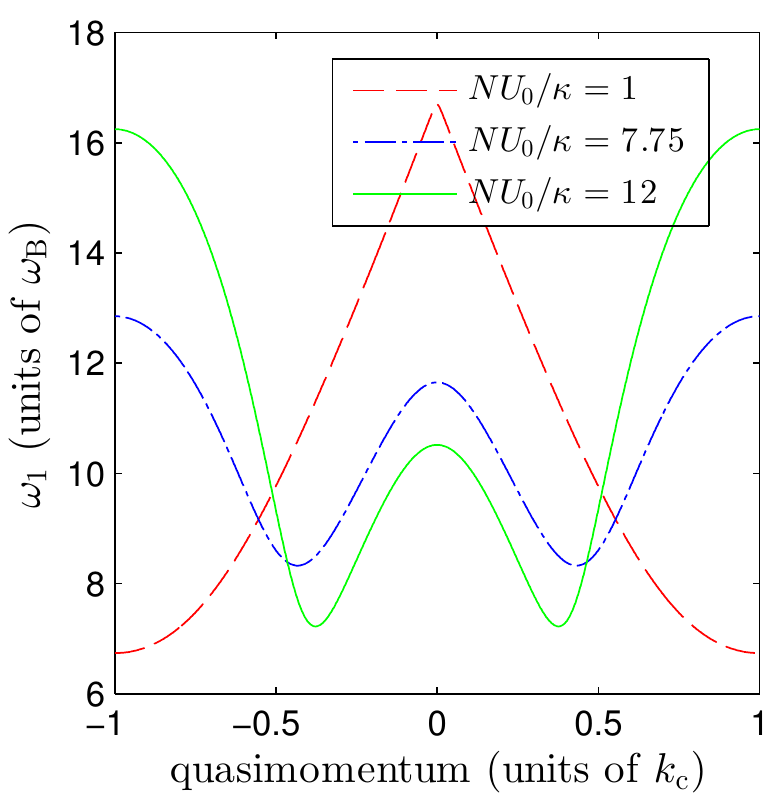}}
  \hfill
 \subfloat[Lowest excitation frequency for all $q$ as a function of
$\beta$]{\label{fig:qpergfnofcoup}
\includegraphics[width=.45\textwidth]{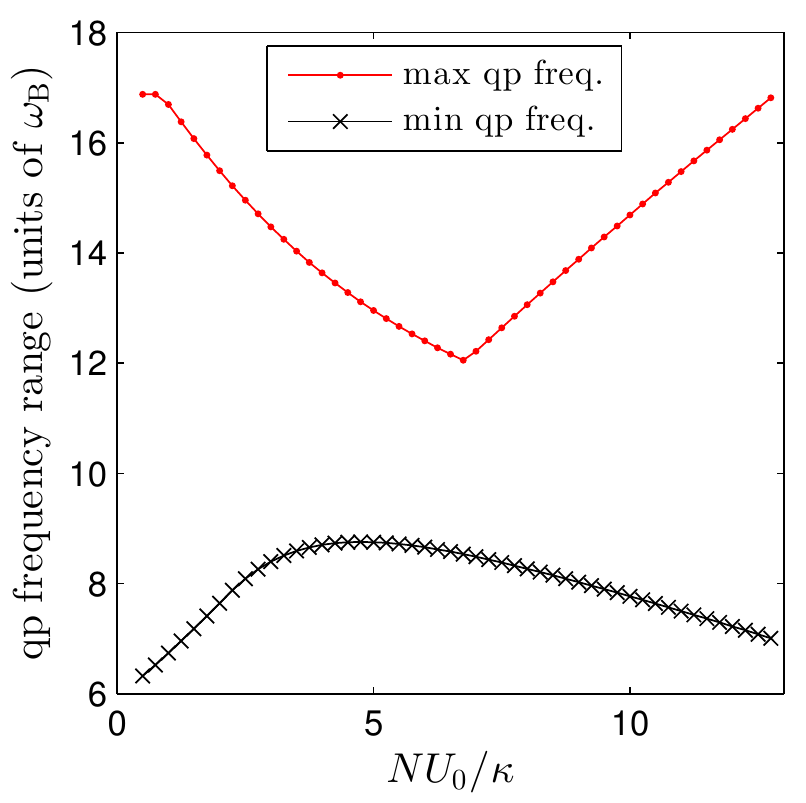}}
\caption{(Color online) Plots of the lowest quasiparticle excitation frequency $\omega_{1}$ about
the adiabatic solution. In (a) this is
given as a function of quasimomentum for three different values of $\beta$: For small $\beta$
(red dashed curve) the minimum of the frequency occurs at $q=\pm 1$, but for larger values of
$\beta$ the minimum shifts in to smaller
values of $q$. Since each quasiparticle excitation has an energy varying with $q$, in (b) we plot
the range of possible excitation
frequencies contained in $\omega_{1}$ for all $q$ as a function of $\beta$. The frequency units are
the Bloch frequency
$\omega_{\mathrm{B}}$.}
\label{fig:snrplots3}
\end{figure*} 

In Figs.\ \ref{fig:snrfnlatdtcorr} and \ref{fig:snrfnNdtcorr}  we show how the SNR depends on other
parameters, namely the lattice depth and total number of atoms. In particular, in
\figref{fig:snrfnlatdtcorr} we plot the SNR as a function of the minimum lattice depth in the cavity
for the coupling value $NU_0/\kappa = 1$. The lattice depth is changed by increasing $\eta$. The red
(dots) curve gives the SNR calculated using only meanfield dynamics plus the effect of shot noise at
the detector and justifies the comment made in Section \ref{sec:mftresults} that for larger lattice
depth the contrast decreases. The SNR calculation including measurement backaction fluctuations is
given by the black (crosses) curve, and has the same qualitative behaviour but is somewhat lower.
\figref{fig:snrfnNdtcorr}  plots the SNR as a function of $N$, where for different values of $N$ we
keep $ NU_0/\kappa =1 $ constant by scaling $U_0$. We also scale the pump strength $\eta$ to
maintain the same intracavity lattice depth $s(t)$ in all the cases. As we pointed out in Section
\ref{sec:mftresults}, this method of scaling the system variables leaves the form of the meanfield
and fluctuation equations unchanged. The only quantitative change is that the meanfield cavity field
solution $\alpha(t)$ is scaled by the same  $\sqrt{r}$ factor as the pumping. This leads to a linear
scaling of the SNR as a function of $N$ (with and without fluctuations) as shown in the plot. It is
interesting to note that the rate of increase is different for the calculation including
fluctuations compared to that without. Clearly there is a gain in the SNR with $N$.

\begin{figure}
\includegraphics[width=0.9 \columnwidth]{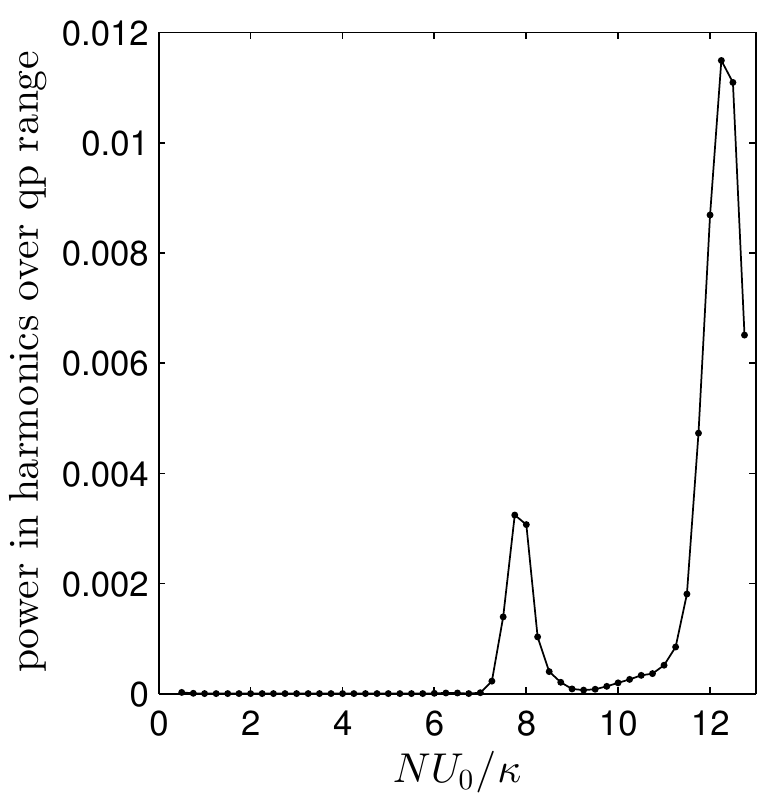}
\caption{The normalized power in the harmonics of $\omega_{\mathrm{B}}$ as calculated from the
Fourier transform of the meanfield solution
(see Fig.\ \ref{fig:mftlatFTatsnrdip}) that lies in the frequency range of the lowest quasiparticle
excitation (see Fig.
\ref{fig:qpergfnofcoup}).}
\label{fig:atomfluctnumres} 
\end{figure}

Finally, we shall explain the physical origin of the complicated series of dips in the SNR when
$\beta>7$ that are seen in \figref{fig:snrfnofcoup1}. Consider the spectrum of quasiparticle
excitations about the adiabatic meanfield solution introduced in Section \ref{sec:excspect}. For the
example shown in \figref{fig:spectrealplotdc1_coup2}, the smallest excitation frequency occurs at
the band edge $q=\pm 1$ and the largest at $q=0$.  As $\beta$ is increased in the usual manner
(holding the minimum lattice depth constant), the $q$-dependence of the quasiparticle spectrum
evolves, as shown for the quasiparticle mode $\omega_{1}$ in \figref{fig:specreseff}. Thus, the
range of frequencies (i.e.\ across the entire Brillouin zone) contained in $\omega_{1}$ also evolves
with $\beta$ and is shown in \figref{fig:qpergfnofcoup}. If the meanfield dynamics happens to
contain any frequencies that fall in this range there is clearly the possibility of a resonance,
exciting quasiparticles and lowering the SNR. This is exactly what happens as can be seen from Fig.\
\ref{fig:atomfluctnumres} which plots the total power in the harmonics of $\omega_{\mathrm{B}}$ that
fall in the frequency range covered by $\omega_{1}$. The two peaks in Fig.\
\ref{fig:atomfluctnumres} at $\beta \approx 8$ and $\beta \approx 12$ coincide exactly with the dips
in \figref{fig:snrfnofcoup1}. Referring back to the inset in Fig.\ \ref{fig:mftlatFTatsnrdip}, which
was deliberately evaluated at $\beta=7.75$ for this very purpose, we can see the part of the
meanfield spectrum that falls in the range spanned by $\omega_{1}$. In the absence of BOs the
quasiparticle excitation $\omega_{1}$ is very narrow, with a width given by the imaginary part
$\gamma_{1}$ evaluated at $q=0$. However, the BO dynamics effectively broadens  the resonance  by
orders of magnitude to that shown in \figref{fig:qpergfnofcoup} and this has a dramatic effect on
the SNR.

\section{Discussion and Conclusions}
\label{sec:conclusion}

In this paper we have extended our previous analysis of Bloch oscillations of ultracold atoms inside
a cavity to include the effects of quantum noise in the electromagnetic field. The quantum noise
originates from the open nature of the cavity and can be interpreted as a form of quantum
measurement backaction because it perturbs the dynamics.  The magnitude of the backaction is
controlled by the dimensionless atom-light coupling parameter $\beta=N U_{0}/\kappa$ and we find
that it can strongly affect the sensitivity of a measurement of the Bloch oscillation frequency
$\omega_{\mathrm{B}}$ and therefore the determination of the magnitude of the external force $F$
driving them.

Our treatment is based upon the coupled Heisenberg equations of motion for the atoms and light which
we linearize about their meanfield solutions, i.e.\ a Bogoliubov level approximation.  We solve the
\emph{time-dependent} meanfield level dynamics exactly and hence coherent effects such as Landau-Zener tunneling between
bands are fully taken into account. A spectral decomposition of the meanfield solution shows that it
is dominated by $\omega_{\mathrm{B}}$ and its first few harmonics, but as $\beta$ is increased
spectral power begins to spread to higher frequencies. 

Quantum noise is introduced via Langevin operators which act as inhomogeneous source terms in the
Heisenberg equations. These terms excite quasiparticles  (quantized excitations with a mixed atom-light character) out of the meanfield. In the standard situation \cite{horak01,gardiner01,szirmai10} where there is no external force then if the system is started off with no quasiparticles their number initially grows in time but eventually saturates due to competition between cooling and heating processes (provided we are in the cooling regime $\Delta_{c}^{\mathrm{eff}}= \Delta_{c}-NU_{0} \langle
\cos^{2}(x)\rangle < 0$ which means that the quasiparticle energy has a negative imaginary part). By contrast, in this work we have found that the presence of an external force, and hence BOs, profoundly changes this behaviour so that following some initial transients the heating rate settles down to a constant value even when we are nominally in the cooling regime. Nevertheless, for the parameter regimes we tested the heating rate was modest and the fraction of the atoms excited out of the coherent meanfield over the lifetime of the simulation was always less than 1\% even for quite strong coupling.

In order to gain some insight into the numerical calculations we used Fermi's golden rule to develop
a semi-analytic model for the heating rate in terms of a simple rate equation for the number of
atomic excitations. In so doing we approximated the cavity light field by a coherent state whose
quantum fluctuations are the same as those of the vacuum. This is a common approximation in cavity
optomechanics but ignores the quantum correlations that build up between the atoms and the light.
Comparing this with the exact numerical results for the number of atomic excitations, we infer that
the field is close to a coherent state for small $\beta$, but differs from it as $\beta$ is
increased, as expected. Furthermore, this comparison allowed us to see the dynamic generation of atom-light correlations.

The above calculations can be applied to the estimation of the signal-to-noise ratio for a
continuous measurement of $\omega_{\mathrm{B}}$. For example, we find that the SNR decreases with
intracavity lattice depth, and increases with the number of atoms. Our principal result, however,
concerns the dependence upon $\beta$. We find that the SNR can be severely reduced due to resonances
between the quasiparticle spectrum and the Bloch oscillating meanfield for certain ranges of
$\beta$. Indeed, the SNR behaviour depicted in Fig.\ \ref{fig:snrplots1} is much more complicated
than that found in the standard example of a quantum limited position measurement of a harmonic
oscillator, e.g.\ the end mirror of a resonant cavity \cite{clerk10}. In that system, the SNR is
determined by the competition between the ``measurement imprecision'' (detector shot noise), which
decreases with increasing measurement strength, and the measurement backaction, which increases with
increasing measurement strength, and correlations between the two can be ignored to a good
approximation. This leads to a smooth curve (see Fig.\ 5 on p.1171 of the review \cite{clerk10})
with a single maximum at the measurement strength where the two effects are equal. This is where the
measurement should be performed for maximum sensitivity. By contrast, in our case we have a cloud of
atoms occupying Bloch states in an optical lattice and thus our system does not correspond very well
to a single harmonic oscillator (except in the limit where the lattice is extremely weak so that the
atoms are predominantly in a state which is uniform in space \cite{brennecke08}, but then
Landau-Zener tunneling will be so severe that the atoms will quickly fall out of the lattice when
the external force $F$ is applied). Add to this the fact that our system is driven by an external
force and so scans through the entire Bloch band in a time-dependent fashion, leading to the
possibility of resonances, and it is not surprising that our resulting SNR in Fig.\
\ref{fig:snrfnofcoup1} does not have a simple maximum as a function of $\beta$. However, we can make
the parameter dependent statement that it seems safest to choose $\beta < 7$ which lies below the
point where the resonances set in (and for $\beta>25$ we find optical bistability which will destroy
the Bloch oscillations \cite{bpvenkatesh11}). The resonances only occur in the calculation when
quantum measurement backaction is included and so provide a salutary example of when the latter is
important. Nevertheless, away from the resonances the SNR for this continuous measurement is large
and is in pretty good agreement with an approximate calculation based upon purely meanfield dynamics
in the cavity plus detector shot noise.

In comparison to previously studied cold atom cavity-QED systems, or even cavity optomechanical
systems, a new feature of our Bloch oscillating system is the time-dependence of the meanfield.
Apart from the resonances discussed above, this also has implications for the computational scheme
we use to calculate the results. For example, all the fluctuation modes should be orthogonal to the
meanfield mode as well as to each other, and hence they also evolve with time. Furthermore, the
two-time correlation functions that are needed to calculate the signal variance that enters the SNR
are not stationary in time, meaning that a large amount of data must be stored. This is especially
true because the Bloch period is roughly three orders of magnitude larger than the quantum
fluctuation timescale $1/\kappa$ and hence the calculation of the SNR over even a few Bloch periods
is quite intensive in the regime where the coherent state approximation breaks down. In  non-cavity
BO experiments it has been shown that coherent dynamics can run for thousands of Bloch periods
\cite{ferrari06}. In a continuous measurement scheme, such as that proposed here, the quantum
measurement backaction reduces the coherence time but unfortunately we have been unable to go much
beyond ten Bloch periods with our numerical computations of the SNR and thereby find this coherence
time for our scheme (we have, however, given an estimate in \cite{bpvenkatesh09} based upon the idea
that the spontaneous emission rate sets the upper limit on coherent dynamics). Nonetheless, our
short time calculations illustrate quantitatively  that it may be advantageous to remain at small
$\beta$ and integrate for longer times. 

\section{Acknowledgements}
This research was funded by the Natural Sciences and Engineering Research Council of Canada. We
thank A. Blais, E. A. Hinds, J. Goldwin, J. Larson, and M. Trupke for discussions related to this
work and D. Nagy for helpful correspondence.

\appendix

\section{Coherent State Approximation}\label{app:shotnoise}

In this appendix we provide details of how the coherent state approximation introduced in
Section \ref{sec:fluctdyn} can be used to derive a simple rate equation for the occupation numbers of atomic fluctuation modes.
Working in the TF, consider the atomic fluctuation operator $ \bthpsip(t)$. Rather than expanding it in plane waves like in Eq.\ (\ref{eq:atomopsplit}), let us instead expand  in the instantaneous eigenbasis $ \nu_k (x,t) $
of the time-dependent meanfield hamiltonian $\bmH(t)$ [given in Eq.\ (\ref{eq:spHtrans})]
\begin{align}
 \bthpsip(x,t) &= \sum_{k}  \nu_k(x,t)  \db_k(t) \label{eq:localbasdecomp}\\
\mbox{where} \quad  \bmH(t)  \nu_k(x,t)  &= E_{k}(t) \nu_k (x,t)  \ . \label{eq:localbasdef}
\end{align}
Substituting the decomposition
\eqnref{eq:localbasdecomp} into the equation of motion \eqnref{eq:fluctatom} we obtain 
\begin{widetext}
\begin{align}
\frac{d \db_j(t)}{dt} &= - i E_j(t) \db_j(t) - \displaystyle \sum_{k} \langle \nu_j(t) \vert
\frac{d}{dt} \vert \nu_k(t) \rangle \db_k(t)
\label{eq:instdecomp}\\ 
&-i \sqrt{N}U_0 \left( \alpha^{*}(t) \dalp(t) + \alpha(t) \dalpd(t) \right) \langle \nu_j(t) \vert
\hP(t) \cos^2(x)\vert \bvphi(t)\rangle \ .
\nonumber
\end{align}
\end{widetext}
We see that the dynamics of the $\db_j(t)$ are coupled amongst themselves: this is obvious from the second term on the right hand side, but also occurs due to the third term as can be seen from \eqnref{eq:lightformalsol}. In order to obtain a description in terms of independent oscillators the contribution from these two terms must vanish, and we will now examine when this happens.

We begin with the second term (with the time derivative) on the right hand side of
\eqnref{eq:instdecomp}. It can be shown that \cite{bohm01}:
\begin{align*}
 \langle \nu_j(t) \vert \frac{d}{dt} \vert \nu_k(t) \rangle \stackrel{j \neq k}{=}
\frac{1}{E_k(t)-E_j(t)} \langle \nu_j(t) \vert \frac{d
\bmH(t)}{dt}\vert \nu_k(t) \rangle
\end{align*}
In general, contributions to the above overlap element are suppressed for states well
separated in energy due to the denominator. Also, we will show below that for $k=j$ the element
vanishes. Hence the dominant contribution comes from adjacent levels i.e.\ $k=j \pm 1$ and is given
by
\begin{align}
 \langle \nu_j(t) \vert \frac{d}{dt} \vert \nu_{j\pm1}(t) \rangle &= \frac{1}{E_{j\pm 1}(t)-E_j(t)}
\langle \nu_j(t) \vert \frac{d \bmH(t)}{dt}\vert \nu_{j \pm 1}(t) \rangle \nonumber\\
&= -2 \frac{\omega_B}{\pi \Delta_{\pm}} \langle \nu_{j} \vert \hat{p} \vert \nu_{j \pm
1} \rangle \label{eq:localbasovlap}
\end{align}
where the second line is obtained by taking a derivative of the instantaneous hamiltonian and
realizing that, due to the opposing relative parity of adjacent states, the term in the overlap integral due to the
potential is zero. The above term can be neglected if the Bloch frequency is
small compared to the energy gap $\Delta_{\pm}$. To proceed further we will assume that this is the case
but in the next appendix we will see that this cannot be guaranteed in general. Specifically,
this approximation is most likely to break down at the times when the quasimomentum comes close to the center or the
edge of the Brillouin zone where there are avoided crossings. Thus, this calculation will be valid
only for short times (since at longer times the system will have repeatedly gone through such
crossings) and/or at parameter regimes where the gaps are large compared to Bloch frequency. 

Coming back to the case when $k =j$ we have $\langle \nu_k (t)\vert \nu_k (t)\rangle=1$ and so
\begin{align*}
 \frac{d}{dt}(\langle \nu_k (t)\vert \nu_k (t)\rangle) = \langle \nu_k(t) \vert \frac{d}{dt} \vert
\nu_k(t) \rangle + h.c. = 0
\end{align*}
i.e.\ the derivative is purely imaginary. For the time dependent hamiltonian
$\bmH(t)$ the potential term $\cos^2(x)$ has an inversion symmetry about $x=0$ and we can always
choose the instantaneous eigenbasis $ \nu_k(x,t) $ to have real coefficients when expanded
over plane waves. As a result the above term goes to zero and the second term in Eq.\  (\ref{eq:instdecomp}) can be
excluded. 

Turning now to the third term on the right hand side of \eqnref{eq:instdecomp}, we can see from
\eqnref{eq:lightformalsol} that it does not couple the different modes $\db_j(t)$ if the light field fluctuations are
independent of the atomic fluctuations i.e.\
\begin{align}
 \dalp(t) \approx \hd(t) \equiv \sqrt{2 \kappa} \int_0^t d \tau e^{-i A(t) (t- \tau)} \hxi(\tau) 
\label{eq:shotnoiseapp}
\end{align}
which is exactly the coherent state approximation. 

Having now seen the conditions under which the fluctuations in the instantaneous eigenmodes of $\bmH(t)$  become independent, let us assume that these conditions are fulfilled so that the fluctuations obey the uncoupled equations of motion
\begin{align}
  \frac{d \db_j}{dt} &= -i E_j \db_j(t) - u_j(t) \hat{\mathcal{F}}(t) \label{eq:couplingtimedep} 
  \end{align}
  where
  \begin{align}
u_j(t) &= i \sqrt{N}U_0 \langle \nu_j(t) \vert \hP(t) \cos^2(x)\vert \bvphi(t)\rangle  \\
\mbox{and} \quad \hat{\mathcal{F}}(t) &= \left( \alpha^{*}(t) \hd(t) + \alpha(t) \hdd(t) \right) . 
\end{align} 
These equations describe the atomic fluctuation dynamics in terms of a collection of
\emph{independent} oscillator modes that are acted upon by the shot noise force
$\hat{\mathcal{F}}(t)$. As described in \cite{marquardt08}, we can now use Fermi's golden rule to
derive a rate equation for each of the oscillator occupation numbers $\delta N_j(t) = \langle
\db_j^{\dagger} \db_j \rangle$
\begin{align}
 \frac{d \langle \delta N_j \rangle}{dt} = \left( \Gamma_{uj} - \Gamma_{dj} \right) \langle \delta
N_j \rangle + \Gamma_{uj}
\label{eq:indoscexceqn}
\end{align}
where the damping and diffusion rates are
\begin{align*}
\Gamma_{uj} =  \vert u_j \vert^2 S_{\mathcal{FF}}(-\omega_j); \quad \Gamma_{dj} = \vert u_j \vert^2
S_{\mathcal{FF}}(\omega_j) \, .
\end{align*}
These depend on the spectral density (power spectrum) of the shot noise force
\begin{align}
S_{\mathcal{FF}}(\omega) = \frac{2\kappa \bar{n}}{(\dceff+\omega)^2+\kappa^2}
\label{eq:shotnoisespec}.
\end{align}
In the above expressions the shot noise spectrum is evaluated at the shifted oscillator frequencies
defined by $\omega_j = E_j(t)-\mu(t)$ with the instantaneous chemical potential $\mu(t) = \langle
\bvphi(t) \vert \bmH(t) \vert \bvphi(t) \rangle$. This shifting helps in removing the slow time
dependence of the couplings $u_j$ (derived from the meanfield Bloch oscillations). Since the damping and diffusion rates for the different oscillators are not the same, it is in general not possible to write down an equation similar in form to
\eqnref{eq:indoscexceqn} for the total $\delta N(t)$, and we have to settle instead for $\delta N(t) =  \sum_j \delta N_j(t)$.

\begin{figure*}
\centering
  \subfloat[Real part of quasiparticle spectrum as a function of time]
{\label{fig:qpbandcrosscoup1}\includegraphics[width=.45\textwidth]{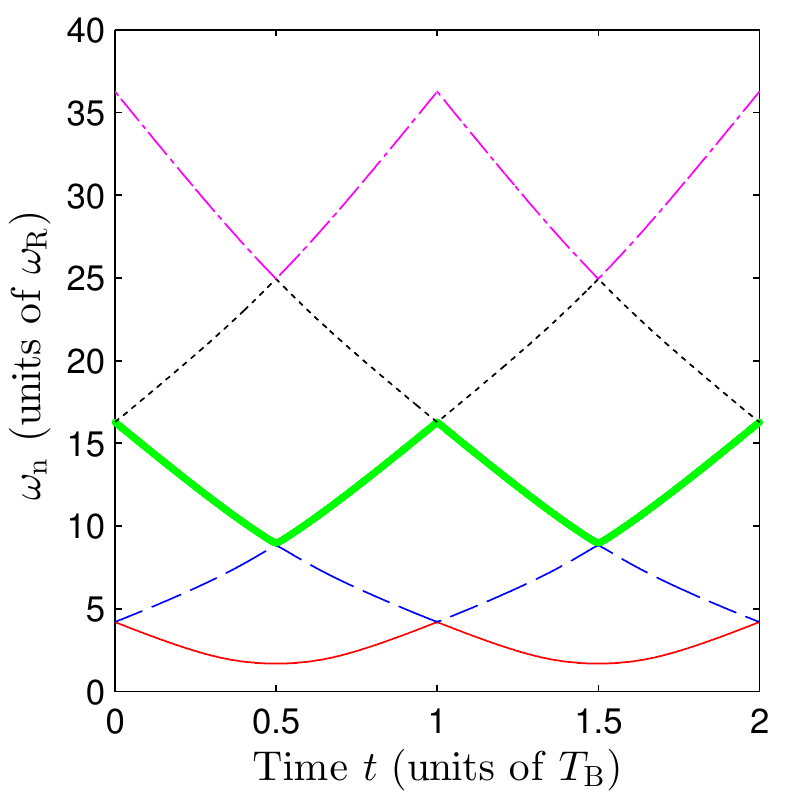}}
  \hfill
 \subfloat[Quasiparticle occupation number as a function of time]
{\label{fig:qpnumbcoup1} \includegraphics[width=.45\textwidth]{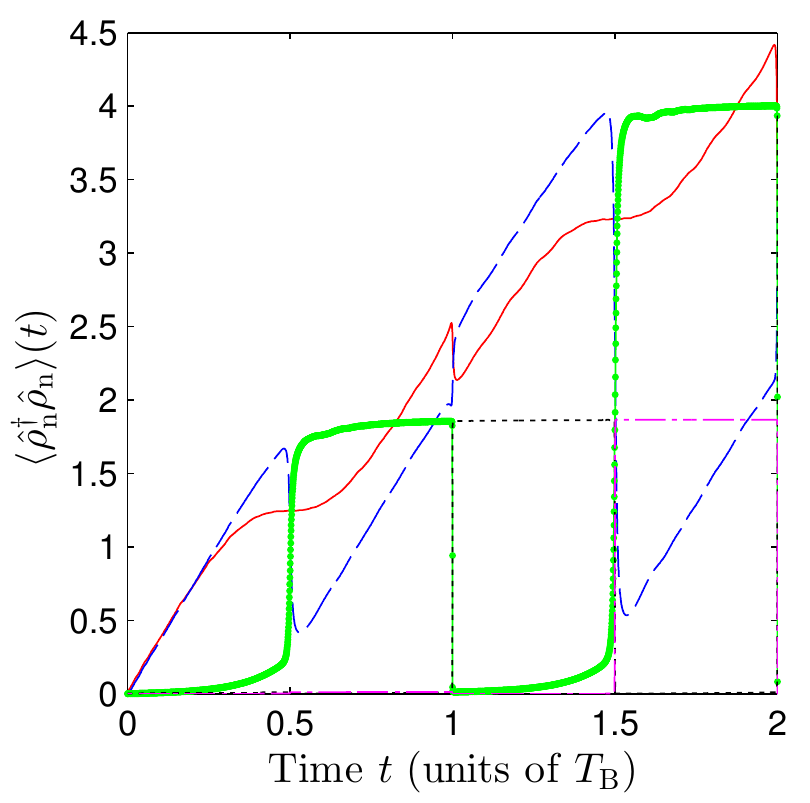}}
\caption{(Color online) Plots of (a) the real part of the quasiparticle energy spectrum and (b) the occupation
number as a function of time. The two plots are colour coded equivalently. For example, the red
(solid) lowest lying level in (a) has occupation number dynamics shown by the red (solid) line in
(b). Since the gaps in the spectrum in (a) are smaller than the Bloch frequency the level
populations are partially exchanged at the avoided crossings: the gaps become smaller higher up in the spectrum and indeed we see that the exchanges between higher lying states are almost complete. The
system parameters are the same as the case with $NU_0/\kappa=1$ in \figref{fig:mftlatplotdc1_changecoup}}
\label{fig:coup1qpoccdyn}
\end{figure*} 

\begin{figure}
\includegraphics[width=0.9 \columnwidth]{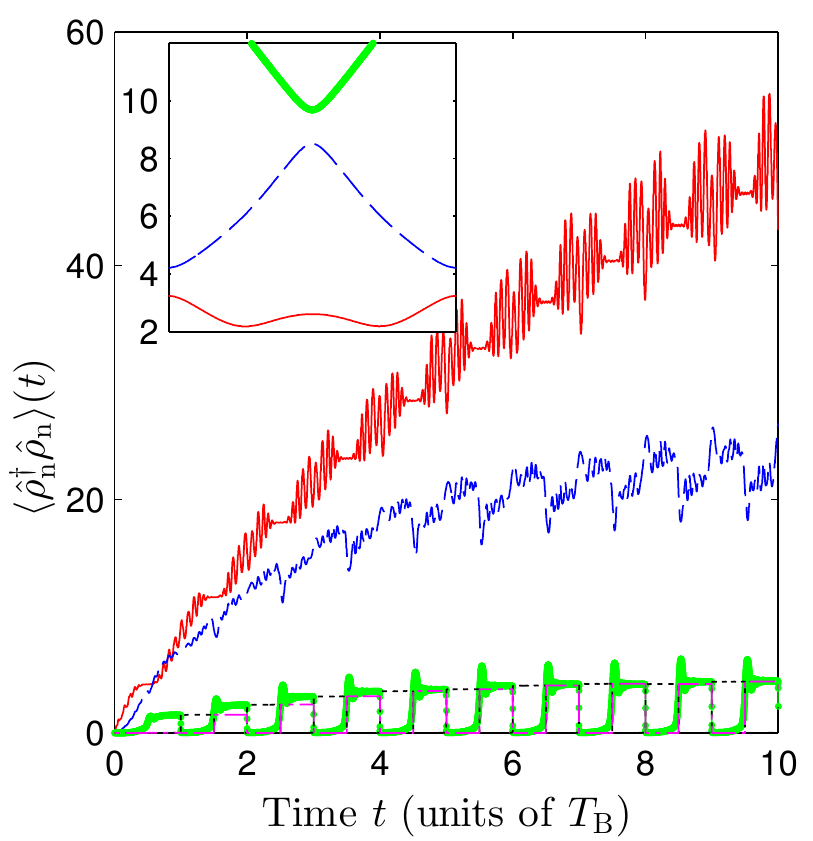}
\caption{Plot of quasiparticle (qp) occupation number as a function of time when
$NU_0/\kappa = 5$. The red (solid) line corresponds to the qp band with the smallest energy, followed by
the blue (dashed), green (dot dashed), black (dotted) and magenta (dash dotted) lines in ascending
order. The inset shows the real part of the qp energy measured in units of $\omega_{\mathrm{R}}$  as a function of time over a single Bloch period $T_{\mathrm{B}}$  for the lowest three bands. Since the
gap between the lowest two bands (red (solid) and blue (dashed) lines) and the rest of the spectrum
is larger than the Bloch frequency, their dynamics is decoupled from the rest.
System parameters are as in \figref{fig:mftlatplotdc1_changecoup}.}
\label{fig:qpnumbcoup5} 
\end{figure}

\begin{figure}
 \includegraphics[width = 0.9\columnwidth]{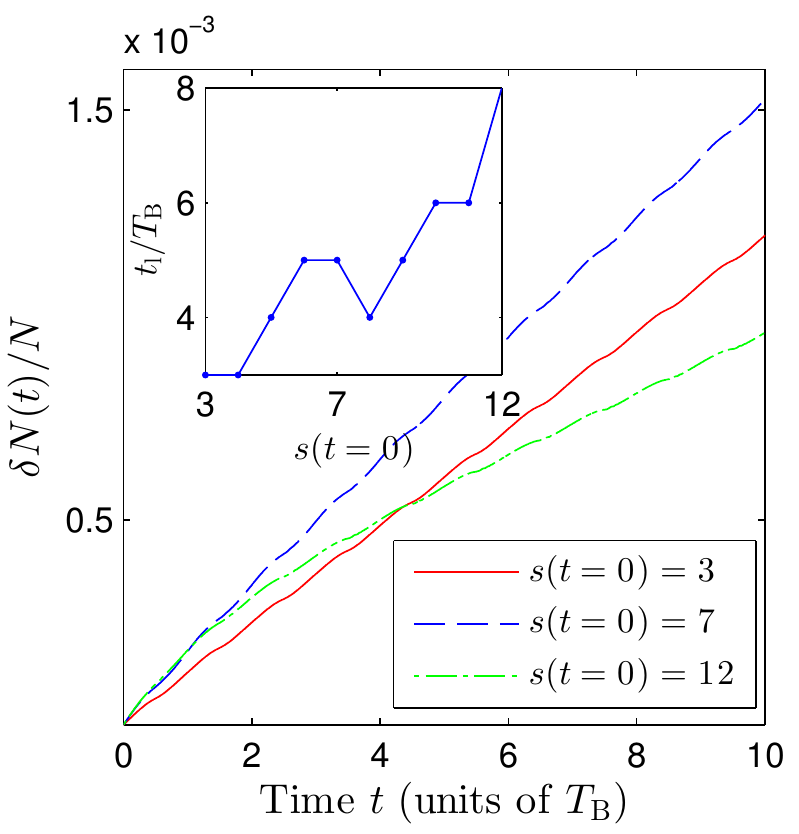}
\caption{Plot of the quasiparticle occupation number as a function of time for different values of
initial lattice depth at the fixed coupling value $\beta = NU_0/\kappa = 1$. The initial lattice
depths $s(t=0)$ are measured in the units of $\omega_{\mathrm{R}}$ and are obtained by setting the pump-strength to $\eta = \{ 44.2, 56.1\} \kappa$ for the blue (dashed) and green (dash dotted) curves, respectively.  The inset plots the time $t_l$ at which the linear
increase in the quasiparticle number is established as a function of the initial lattice depth.  }
\label{fig:linincdeplat} 
\end{figure}
\section{Absence of cavity cooling in the presence of Bloch oscillations} \label{app:longtimebehav}

In this appendix we analyze the long-time behaviour of the number of atomic fluctuations $\delta N$. We do this in order to understand the apparent absence of a cavity cooling effect in the results shown in \figref{fig:fluctphatnuma}. In the standard case where there is no external force \cite{szirmai09,szirmai10}, cavity cooling occurs when the effective detuning $\dceff \equiv \Delta_{\mathrm{c}} - NU_0 \langle \cos^2(x) \rangle$ is negative. This ensures that the quasiparticle energies have a negative imaginary part $\gamma_{n}<0$ which implies dynamical stability as explained in Section \ref{sec:excspect}. Under these circumstances $\delta N$ reaches a steady state and the heating rate vanishes as shown in the inset in Fig.\ \ref{fig:fluctphatnuma}. This is, however, not what we see in the presence of an external force as shown in the main body of  \figref{fig:fluctphatnuma} where the heating rate settles down to a constant nonzero value. The external force must therefore disrupt the cooling mechanism, and in this appendix we shall see that indeed the periodic driving due to the BOs drives the quasiparticles to higher energy states thereby heating the system.

The heating rate is given by the change in the occupation numbers of the various quasiparticle states as
a function of time.  These states are nothing but the instantaneous eigenvectors of the fluctuation matrix $\bM(t)$ introduced in Section \ref{sec:excspect}  \cite{szirmai09}
\begin{align}
 \bM(t) \ r^{(n)}(t) &= (\omega_n(t) + i \gamma_n(t)) \ r^{(n)}(t) \ , \label{eq:qpdefn}
\end{align}
and have a mixed atom-photon character.
However, the fluctuation matrix $\bM(t)$ is non-normal and so its left and right
eigenvectors are not the same. The left eigenvectors $l^{(n)}$ are defined as 
\begin{align}
 \bM^{\dagger}(t) \ l^{(n)}(t) &= (\omega_n(t) - i \gamma_n(t)) \ l^{(n)}(t).
\end{align}
The left eigenvectors can be used to define the quasiparticle mode operator $\hat{\rho}_n(t)$ corresponding to the
$n$th mode as
\begin{align}
 \hat{\rho}_n(t) \equiv \left(l^{(n)}(t), \hat{R}(t)\right)
\end{align}
where the bracket on the right in the above equation denotes a scalar product and $\hat{R}(t) = \left(
\dalp(t) \, \dalpd(t) \, \bthpsip(t) \, \bthpsip^{\dagger}(t) \right )^{T}$ is the fluctuation operator in the basis of atoms and photons [see \eqnref{eq:mateqn}]. We therefore see that the required quasiparticle occupation numbers  $\langle \rho_n^{\dagger} \rho_n \rangle (t)$ as
a function of time can easily be computed from the numerical solution of the covariance matrix
$\bC(t)$ [\eqnref{eq:covmatsol}] once the eigenvectors $l^{(n)}(t)$ are obtained. Before we look at
the results, we should first comment on the relation between the quasiparticle occupation number
and the atomic fluctuation number $\delta N(t)$. As mentioned in Sec.\ \ref{sec:excspect},
quasiparticle modes come in three types and the most relevant ones are the hybridized atom-light
modes which have the strongest atom-light coupling and tend to lie lowest in the spectrum. Since the hybridized modes contain both atomic and light components, their occupation number
is not exactly equal to the atomic fluctuation occupation number.
Nonetheless, in this system the atom-light entanglement is not very large \cite{szirmai10} and the
total quasiparticle occupation number closely tracks the atomic fluctuation number (as we have
verified). Moreover, to establish a connection with the
calculation in Appendix \ref{app:shotnoise}, we note that for small $NU_0/\kappa$ the atomic part
of the hybridised quasiparticle modes are very close to the higher band eigenstates of the
instantaneous meanfield hamiltonian $\bmH(t)$. Thus, the mode occupation of the oscillators in
Appendix \ref{app:shotnoise} can be roughly mapped to the quasiparticle occupation numbers here.

In order to understand the occupation number dynamics, consider first the real part of the
quasiparticle spectrum plotted in \figref{fig:qpbandcrosscoup1} as a function of time for $NU_0/\kappa = 1$.  The quasiparticle energy bands have avoided crossings every half Bloch period which alternate between being with the band above and below. On the scale of the plot, the gaps at the crossings are not discernible but for the present parameters it turns out that even the gap between the lowest two bands is smaller than the Bloch frequency (recall that in this paper we have set $\omega_{\mathrm{B}}=0.25 \omega_{\mathrm{R}}$) and the magnitude of the gaps gets smaller as we go higher up in the spectrum. During the course of Bloch oscillations these avoided crossings are repeatedly traversed at the Bloch frequency and consequently the occupation number dynamics at the avoided crossings are increasingly non-adiabatic as we go up in the spectrum due to Landau-Zener transitions. For example, in \figref{fig:qpbandcrosscoup1} at $t = T_B/2$ the green (dot-dashed) curve of the third band approaches the blue (dashed) curve of the second band and as a result the populations of the two levels are almost completely exchanged as can be seen at the corresponding time in \figref{fig:qpnumbcoup1}. We therefore have the following picture: the occupation number of a given quasiparticle band increases either by direct scattering out of the meanfield due to quantum noise or by upcoming quasiparticles from the immediately lower band by a Landau-Zener transition. The occupation decreases due to quasiparticles scattering back into the meanfield [the hermitian conjugate term to the excitation processes in Eqns. (\ref{eq:fluctphot}) and (\ref{eq:fluctatom})], or due to the finite lifetime of quasiparticles associated with cavity decay at rate $\kappa$ as described by the $A(t)$ term in Eq.\ (\ref{eq:fluctphot}), or due to Landau-Zener transitions to the next higher band.  This has to be contrasted with the dynamics without BOs where the quasimomentum is fixed at $q=0$ and the fluctuations occupy a stationary quasiparticle ladder. Without Landau-Zener transitions there is no directed transport of quasiparticles up the ladder and cooling effects, due to the finite quasiparticle lifetime $1/\gamma_{n}$, have time to act.

In \figref{fig:fluctphatnuma} notice that the linear behaviour is established at  later
times for larger $NU_0/\kappa$. In order to understand why this happens we explore the quasiparticle
number dynamics for $NU_0/\kappa = 5$ in \figref{fig:qpnumbcoup5}, i.e.\ a factor of 5 greater than in Figs.\ \ref{fig:qpbandcrosscoup1} and \ref{fig:qpnumbcoup1}. From the inset we can
immediately see that the two lowest quasiparticle bands are well isolated (by more than
$\omega_B$) from the rest of the ladder. As a result, the occupation numbers in these modes evolves 
in an adiabatic manner, in contrast to the situation for $NU_0/\kappa = 1$. In fact, over the times plotted in \figref{fig:qpnumbcoup5}, the blue (dashed) band reaches a steady average occupation number. But the higher quasiparticle energy
levels represented, for instance, by the green (dotted) and
black (dot dashed) lines have smaller gaps and behave akin to \figref{fig:qpnumbcoup1} because they are rapidly emptied by Landau-Zener transitions.  Another relevant observation comes from \figref{fig:fluctphatnumb}, where we see that the fluctuation photon
number reaches its quasi-steady state around the same time as the atomic fluctuation number begins to exhibit
\emph{linear} growth. This can be understood now in the light of the above discussion since the lowest
quasiparticle modes are coupled most strongly to the light field. The red (solid) band in the inset
of \figref{fig:qpnumbcoup5} has two minima and demonstrates how for larger $NU_0/\kappa$ the
quasiparticle bands can be strongly modified from the single particle (linear) band structure.

We conclude this appendix by examining another way to control the band gaps in the quasiparticle
spectrum and as a result the time taken for the linear increase behaviour (denoted by $t_l$
henceforth) to set in. In \figref{fig:linincdeplat} we plot the atomic fluctuation number as a
function of time for $\beta=NU_0/\kappa = 1$ and three different initial meanfield lattice depth values
that are set by the pump strength. Since the initial atomic state has $q=0$, the initial lattice
depth is the minimum lattice depth over the Bloch period. Furthermore, we are at relatively small $\beta$,
and so the linear band picture holds good and one can anticipate that $t_l$ increases with lattice
depths due to the widening of band gaps. In the inset in \figref{fig:linincdeplat} we plot
$t_l$ as a function of the initial lattice depth for a range of values at $NU_0/\kappa = 1$. As
expected, we see a general trend of increasing $t_l$ for larger lattice depths. We have identified
$t_l$ from the numerical simulation for atomic fluctuation number by requiring that the average change in
the rate of increase of $\delta N(t)$ over a Bloch period converge to three significant figures.

\section{Two-time correlation calculation} \label{app:corrcalc}

When the intracavity light field is written as $\ha(t) = \alpha(t)+\dalp(t)$, the signal variance is
given by:
\begin{widetext}
\begin{align}
& \langle \Delta \hat{N}^2(\omega,T) \rangle =  \kappa \left [ \int_0^{T}  \cos^2(\omega t) \left(
\vert \alpha(t) \vert^2 +  \langle
\dalp^{\dagger}
\dalp(t) \rangle \right)dt \right] + 2 \kappa^2 \Re \left [ \int_0^{T} dt_1 dt_2 \cos(\omega t_1)
\cos( \omega t_2)\alpha(t_1) \alpha(t_2)
\langle \dalpd(t_1) \dalpd(t_2)
\rangle \right ] \nonumber\\
&+ \kappa^2  \left [ \int_0^{T} dt_1 dt_2 \cos(\omega t_1) \cos( \omega t_2)\alpha^{*}(t_1)
\alpha(t_2) \langle \dalp(t_1) \dalpd(t_2)
\rangle 
+ \int_0^{T} dt_1 dt_2 \alpha(t_1) \alpha^{*}(t_2) \cos(\omega t_1) \cos( \omega t_2) \langle
\dalpd(t_1) \dalp(t_2) \rangle \right] 
\nonumber \\
&-2 (\kappa)^{3/2} \Re \left[ \int_0^{T} dt_1 dt_2 \alpha^{*}(t_1) \alpha(t_2) \cos(\omega t_1)
\cos( \omega t_2)\langle \hxi_b(t_1)
\dalpd(t_2) \rangle \theta(t_2-t_1)\right]. \label{eq:varoutphotds}
\end{align}
\end{widetext}
In this appendix we provide details of how we numerically compute the signal variance (and hence
the SNR). The important extra computational step compared to the covariance matrix calculation in
\eqnref{eq:essentialmatevol} is the evaluation of the two time correlations such as $\langle
\dalpd(t_1) \dalpd(t_2) \rangle$. In the vector notation for the fluctuations, the two time
correlations are elements of the correlation matrix $\bL(t_1,t_2) = \langle \hat{R}(t_1)
\hR^{T}(t_2)\rangle$. The time evolution for the correlation matrix is given by:
\begin{align}
 i \frac{d}{dt} \bL(t,t_0)= \bM(t) \bL(t,t_0) + i \langle \hZ(t) \hR^T (t_0) \rangle.
\label{eq:correvol} 
\end{align}
Let us consider the case when $t>t_0$. Then the last term in above equation gives a correlation
between the Langevin operators at some future time $t$ and the system fluctuation operators at
$t_0$. Due to the delta-correlated nature of the Langevin noise this term will be zero. This means
that \eqnref{eq:correvol} becomes homogeneous and we can solve it with the initial condition at
$t=t_0$, $\bL(t_0,t_0) = \bC(t_0)$. Also note that the time evolution operator for the numerical
evolution in \eqnref{eq:correvol} is same as the one for the covariance matrix [denoted by $\bG(t)$
in \eqnref{eq:essentialmatevol}], which is an expression of the quantum regression theorem
\cite{walls08}. A separate computation for $t<t_0$ is not needed since they are related to the
elements of $\bL(t,t_0)$  with $t>t_0$ by complex conjugation. For example:
\begin{align*}
 \langle \dalpd(t_0) \dalp(t) \rangle = \langle \dalp(t) \dalpd(t_0) \rangle^{*}.
\end{align*}
We can evaluate the correlation $\langle \hxi_b(t_1) \dalpd(t_2) \rangle$ using a similar approach
as above for the time evolution of the vector $\langle \hxi_b(t_1) \hR(t_2) \rangle$. In this case
the initial condition for the evolution is $\langle \hxi_b(t_0) \dalpd(t_0) \rangle =
\sqrt{\kappa}/2$. Since the evolution operators for the correlation matrix and covariance matrix
evolution are the same the calculation can be performed without additional computational cost. The
main difficulty in computing the signal variance arises from the fact that the two time correlation
functions are not stationary. As a result, in order to evaluate the integrals in
\eqnref{eq:varoutphotds} the correlation matrix needs to be computed for all values of
$0<t_1,t_2<T$. This is the memory intensive step in the computation and we simplify the situation by
performing the correlation matrix computation over a coarser grid than the one used in the numerical
solution of \eqnref{eq:essentialmatevol}. This is justified since we find typically the correlation
matrix elements do not change significantly over the very short time steps chosen in the solution of
\eqnref{eq:essentialmatevol}. Moreover, for the results presented in Section \ref{sec:snrresults},
we have taken care to check that the numerical solutions converge to a value independent of the size
of the coarse grid. The necessity of evaluating two-time correlators over a two dimensional time
grid is the main limiting factor to the maximum integration time for the SNR calculations. Another
point to bear in mind is that for $\beta$ values larger than the ones that we have presented
here we have found that the size of the coarse grid needs to be essentially matched with the size of the
finer computational grid over which \eqnref{eq:essentialmatevol} is solved. As a result the
calculation for strong coupling becomes very memory intensive indeed.

\end{document}